\documentclass[review,3p,authoryear]{elsarticle}

\usepackage{natbib}
\usepackage{tikz}
\usepackage[hidelinks]{hyperref}
\usepackage{multirow}
\usepackage{epstopdf}
\usepackage{mathtools}
\usepackage{amsmath,amssymb}
\usepackage{amsthm}
\usepackage{amsfonts}
\usepackage{marvosym}
\usepackage{color}
\usepackage{bbm}
\usepackage{booktabs}
\usepackage{rotating}
\usepackage{multicol}
\usepackage{graphicx}
\usepackage{bigints}
\usepackage{subfig}
\usepackage{pdflscape}
\usepackage{longtable}
\usepackage{relsize}

\DeclareMathSizes{11}{11}{5}{3.5}

\biboptions{round}

\def\balpha{\mbox{\boldmath $\alpha$}}
\def\bbeta{\mbox{\boldmath $\beta$}}

\def\bvarepsilon{\mbox{\boldmath $\varepsilon$}}

\def\bu{\mathbf{u}} 

\def\by{\mathbf{y}} 

\def\0{\mbox{\bf{0}}}

\def\bI{\mathbf{I}}

\def\bX{\mathbf{X}}

\def\bv{\mathbf{v}}%
\def\sb{\mathsf{b}} 
 
\def\si{\mathsf{i}}

\def\sp{\mathsf{p}}

\def\sL{\mathsf{L}}

\def\sW{\mathsf{W}}

\def\bu{\mathbf{u}}

\def\by{\mathbf{y}}

\def\0{\mbox{\bf{0}}}

\def\bI{\mathbf{I}}

\def\bX{\mathbf{X}}

\def\bv{\mathbf{v}}%
\def\diag{\mbox{diag}}

\newcommand{\suchthat}{\;\ifnum\currentgrouptype=16 \middle\fi|\;}

\newcommand\x{\mathbf x}

\theoremstyle{plain}

\theoremstyle{definition}

\theoremstyle{remark}

\begin{document}

\begin{frontmatter}



\title{\LARGE The Effect of Weather Conditions on Fertilizer Applications: \\ A Spatial Dynamic Panel Data Analysis}

\author[anna]{Anna Gloria Bill\'{e}}
\author[marco]{Marco Rogna}

\address[anna]{Department of Statistical Sciences, University of Padua, Italy}
\address[marco]{Faculty of Economics and Management, Free Univerisity of Bolzano--Bozen, Italy}

\cortext[corresp]{Anna Gloria Bill\'{e}. E--mail: annagloria.bille@unipd.it. Marco Rogna. E--mail: Marco.Rogna@unibz.it.}
\begin{abstract}
Given the extreme dependence of agriculture on weather conditions, this paper analyses the effect of climatic variations on this economic sector, by considering both a huge dataset and a flexible spatio--temporal model specification. In particular, we study the response of N--fertilizer application to abnormal weather conditions, while accounting for other relevant control variables. The dataset consists of gridded data spanning over 21 years (1993--2013), while the methodological strategy makes use of a spatial dynamic panel data (SDPD) model that accounts for both space and time fixed effects, besides dealing with both space and time dependences. Time--invariant short and long term effects, as well as time--varying marginal effects are also properly defined, revealing interesting results on the impact of both GDP and weather conditions on fertilizer utilizations. The analysis considers four macro--regions -- Europe, South America, South--East Asia and Africa --  to allow for comparisons among different socio--economic societies. In addition to finding both spatial (in the form of {\em knowledge spillover effects}) and temporal dependences as well as a good support for the existence of an environmental Kuznets curve for fertilizer application, the paper shows peculiar responses of N--fertilization to deviations from normal weather conditions of moisture for each selected region, calling for ad hoc policy interventions.   
\end{abstract}
\end{frontmatter}
%


%
\section{Introduction}
\label{sec:intro}
Agriculture is one of the sectors  most dependent on weather conditions, being its output strongly affected by them. With the 45\% of the world population living in rural areas \citep{WBrurpop17} and the 26\% of the working population being employed in the primary sector in 2017 \citep{WBruremp17}, it is therefore highly interesting to analyse the effect of weather conditions on agricultural production. This is even more true for developing countries where the percentage of people deriving their livelihood from the primary sector grows tremendously: the percentage of working population being employed in agriculture in 2017 is equal to 32 in low and middle income countries, rising to 68 when considering only low income ones \citep{WBruremp17}.

In a world characterized by global warming and more frequent weather extremes, see e.g. \cite{stott04}, \cite{pall11} and \cite{fischer15}, a strong focus has been placed in understanding the effect that such phenomena have on agricultural output. For instance, \citet{lesk16} evaluated the impact of weather extremes on global crop production, finding a significant negative impact (minus 9--10\% for cereals). \citet{rosenzweig02} limited their analysis to U.S. and to floods and excesses of rain, whereas \citet{zipper16} focused on droughts in the same area. The response to and the adopted strategies to cope with weather extremes is another well established field of investigation. \citet{iizumi15}, among others, analysed the changes in cropping area and intensity due to weather shocks at global level, whereas \citet{morton15} investigated the strategies adopted by Upper Midwest U.S. farmers to cope with a prolonged time of excesses of rain. Several papers examined the farmers' response to droughts in vulnerable areas, e.g. \citet{campbell99} in Kenya, \citet{fisher15} in Sub--Saharan Africa and \citet{hossain16} in Bangladesh.

This paper can be inserted into the above mentioned literature strand, featuring several peculiarities. Its aim is to explore the farmers' response, in terms of fertilizer application, to abnormal weather conditions -- dryness and wetness --, but without focusing exclusively on extreme events. Furthermore, it also focuses on very different world macro--regions rather than specifically taking into account a single geographical area, to account for potential heterogeneity in the N--fertilizer use. Given the importance of fertilizers in enhancing agricultural output, this appears to be a relevant research topic. Moreover, the effect of abnormal weather conditions on fertilizer application is far from easily predictable. Our focus encompasses both the immediate response to dryness and wetness conditions and, particularly, the lagged response. In fact, from a policy perspective, it is important to understand which are the lasting consequences of these weather phenomena. 

If a drought or an excess of rain is happening or is foreseen to happen at the time of fertilizer application, the immediate response of a farmer could be to increase the amount of fertilizer to counterbalance the likely output loss caused by the adverse weather conditions or she could opt to reduce it to avoid wasting input expenditures. Moreover, the response could vary on the base of the intensity of the phenomenon, but also on the base of its nature. According to \citet{purcell04}, the importance of N--fertilization on yield outcomes is more crucial during a drought period than under normal conditions. Regarding excesses of rain, \citet{vlek86} highlighted through several experimental rice plots in South--East Asia that the N--fertilizer uptake efficacy of the crop was dramatically reduced in flooded plots. Although both directions of departure from normal weather conditions seem to have the effect of calling for an increase in fertilization, nothing is said on their relative magnitude. Furthermore, both studies are focused on a specific area and a single crop, therefore it is not granted this to hold in general. 

The lagged response, as to say the application of fertilizer in a time period subsequent the happening of a weather shock, seems to be more easily predictable. Since a significant departure from expected levels of wetness is associated with a reduction in productivity, this implies a lower nutrients uptake by plants. This further causes a lower need of fertilization in the subsequent year\footnote{\label{note:note1}See, for example, the following document from the University of Wisconsin Extension Service at \texttt{http://nasdonline.org/static\_content/documents/1987/d001465.pdf}.}. However, although heavy rains may damage crops as much as drought, they could also cause a run--off of fertilizer from the soil, thus nullifying this argument. The diffused lack of agricultural insurance schemes and the scarcity of farmers financial reserves in developing countries may translate into an obliged reduction of fertilizer application, due to the impossibility of farmers to purchase this input after an income shock \citep{barrett07}. In economically vulnerable rural areas, strong deviations from normal weather conditions may ignite such shock. In less precarious regions, however, it cannot be excluded that farmers simply keep applying the usual amount of fertilizer for the inability to estimate the quantity remained in the soil or even increase its application in the attempt to recover past losses. As just seen, also the lagged effect of abnormal weather conditions on fertilization is rather unclear. Different contrasting hypotheses look theoretically sounding, and the geographical area under investigation may be an important factor in determining which of them holds. An empirical investigation is therefore necessary to unravel this query.

In order to shed light on these hypotheses, covering an almost global dimension, we use spatio--temporal data on the yearly amount of applied synthetic nitrogen (N) fertilizer taken from \citet{lu17}. The time dimension is of 21 years, while the considered areas are Europe (i.e. countries affected by the Common Agricultural Policy, CAP), South America, South--East Asia and Africa, to cover a very broad spectrum of climatic and socio--economic conditions. To deal with the spatio--temporal dimension of our data, we make use of a spatial dynamic panel data model (SDPD), with both individual--specific and time--specific fixed effects to account for potential unit--specific unobserved heterogeneity. Looking at the recent literature on spatial econometrics, see e.g. \cite{elhorst2014spatial} and references therein, although the theoretical issues are quite advanced, see e.g. \cite{shi2017spatial} 
among others, to the best of our knowledge only few empirical applications within agricultural economics and/or climate change have recently used these model specifications: see, for some examples, the interesting review of \citet{baylis2011spatial}. In particular, we are the first ones who make use of the dynamic version of spatial panel data models with spatio--temporal agricultural data. Moreover, because of the nonlinearity in variables in our model specification, we properly define the marginal effects distinguishing between time--varying and time--invariant impacts.  

Our main results confirm a rather strong geographical variability for the effect of abnormal weather conditions on fertilizer application. Whereas in Europe and in South America the response to past abnormal weather conditions seems to follow optimal agronomic practices, in South East Asia and in Africa weather conditions seem to be detached from the farmers' choice of fertilizing. This may be caused by the inability of local farmers to responsively adapt their behaviour to external shocks for lack of either technical skills or means. In these last two regions it is also observed a potential problem of poverty traps since the lagged per capita GDP is a strong determinant of the amount of nutrient present in the soil. Furthermore, a modest evidence for the existence of an environmental Kuznets curve for N--fertilizer is found. Finally, spatial (contemporaneous) and temporal dependence effects as well as spatio--temporal effects heavily contribute to explain N--fertilizer utilization, with only slight differences among the macro--regions. These effects, beyond testifying similar agronomic practices originating from common agricultural conditions (e.g. common soil conditions and grown crops) suggest the existence of knowledge spillovers among farmers. 

The rest of the paper is organized as follows. Section \ref{sec:review} provides a brief literature review on related research works. Section \ref{sec:data} describes the data used. In Section \ref{sec:model} we explain the choice of the model specification, the hypothesis test strategy and we derive appropriate marginal effects to correctly interpret the results. In Section \ref{sec:results} we provide and discuss the results. Finally, Section \ref{sec:conclusion} concludes. 
\section{Literature Review}
\label{sec:review}
In this Section we briefly review the main contributions on the relation between weather extremes and the use of fertilization, distinguishing among different target societies and/or the methodology used.

The impact of weather conditions on agriculture is a largely investigated theme, particularly when considering its extreme forms. \citet{pandey07} dedicate a whole book to examine the effects of droughts on South--East Asian rice farmers and to analyse their coping strategies, evidencing how these lasts are often insufficient to prevent rural households to fall into poverty after severe droughts. \citet{devereux07} develops an analytic framework for understanding the impact of droughts and floods on food security and, based on a weather--induced food crises in Malawi, he uses it to derive policy prescriptions to avoid famine to occur. As mentioned in the introduction, the concerns about climate change have further boosted this literature since droughts, but other weather extremes too, are expected to intensify. \citet{lobell07}, for example, estimate the effect of rising temperatures on the yields of the six most widely grown crops, finding significant negative effects for most of them. The present paper, however, strongly differs from these works in two dimensions. Methodologically, no one of the mentioned works adopts a spatial dynamic econometric approach. With regard to the content, the present paper specifically focuses on the relation between fertilization and weather conditions, rather than examining a broad spectrum of responses to weather shocks. Furthermore, our focus is not limited to extremes weather events, but rather on the whole spectrum of abnormal conditions.

\citet{ding09} share a similar focus on the response to weather variations of a very precise component of agricultural production: tillage practices. Besides the narrowing of the attention to a precise element, they also consider panel data as well as spatial correlation in the error terms, whereas previous works on soil conservation practices mainly used cross sectional data, see e.g. \citet{soule00}. Spatial autocorrelation is then inherent in several agricultural applications. In addition to the obvious difference regarding the outcome of interest -- fertilization rather than tillage -- our model directly considers a spatial process into the fertilization use as well as two dynamic components that lack in \citet{ding09} and a much broader territorial coverage. 

The relation between fertilization and weather conditions, instead, is a topic rather overlooked by the economic literature being confined to the agro-biological domain. \citet{hartmann11} consider the environmental aspect of this interaction, analysing how N--fertilization combined with drought can modify the ability of the soil to store atmospheric methane. \citet{van16} examine the concentration of nitrate in some U.S. Mid--West rivers in a rainy period after a drought, finding a significant increase. This testifies how drought periods effectively leave a high amount of N--fertilizer in the soil. \citet{purcell04} study soy--bean $N_2$ fixation and crop yield response to drought. Among their results, it worth to mention how the yield response to N--fertilization is higher under drought conditions (plus 15--25\%) than in an optimally watered situation (plus 12--15\%).

Given the importance of fertilizers in agricultural production, several studies have focused on estimating the determinants of their spatial diffusion. \citet{potter10} analyse the pattern of both inorganic and organic fertilizer application at world level, whereas \citet{toth14} map the levels of phosphorus (P) present in soil in the European Union. However, a study that relates the application of fertilizer with present and past weather conditions seems, to the best of our knowledge, to lack. This paper tries to fill this gap, by making use of a worldwide database of spatial data. Although this may entail to sacrifice precision compared to using farm specific data derived from surveys, it allows to offer a global view on the topic and to base the analysis on a rather large time span. Moreover, we specify one of the most recent spatio--temporal models which is able to deal with both spatial and temporal dependence structures, as well as a form of spatial and temporal heterogeneity. 
\section{Data description}
\label{sec:data}
In this Section we briefly introduce a description of the data used in our empirical analysis. The dependent variable is the amount of fertilizer applied on a given portion of agricultural land. Relying on data related to quantities at national level would lower the precision of the estimation, since weather extremes may interest only some portions of a country territory and, therefore, their effect may be masked by aggregate data. 

Thanks to a recent dataset made available by \citet{lu17}, it is now possible to overcome this problem. Indeed, the dataset provides global time series gridded data of annual synthetic nitrogen (N) fertilizer application\footnote{The unit of measure of this quantity is grams per square meter per year.} with a resolution of $0.5^{\circ} \times 0.5^{\circ}$ latitude--longitude for the period 1961--2013. Given the high number of available data and the consequent computational burden, and supposing a decreasing quality of data for periods more distant in time, we decided to curtail the considered years from 1992 onward.

\begin{figure}[ht]	
	\centering
	\includegraphics[width=0.8\textwidth,height=0.8\textwidth]{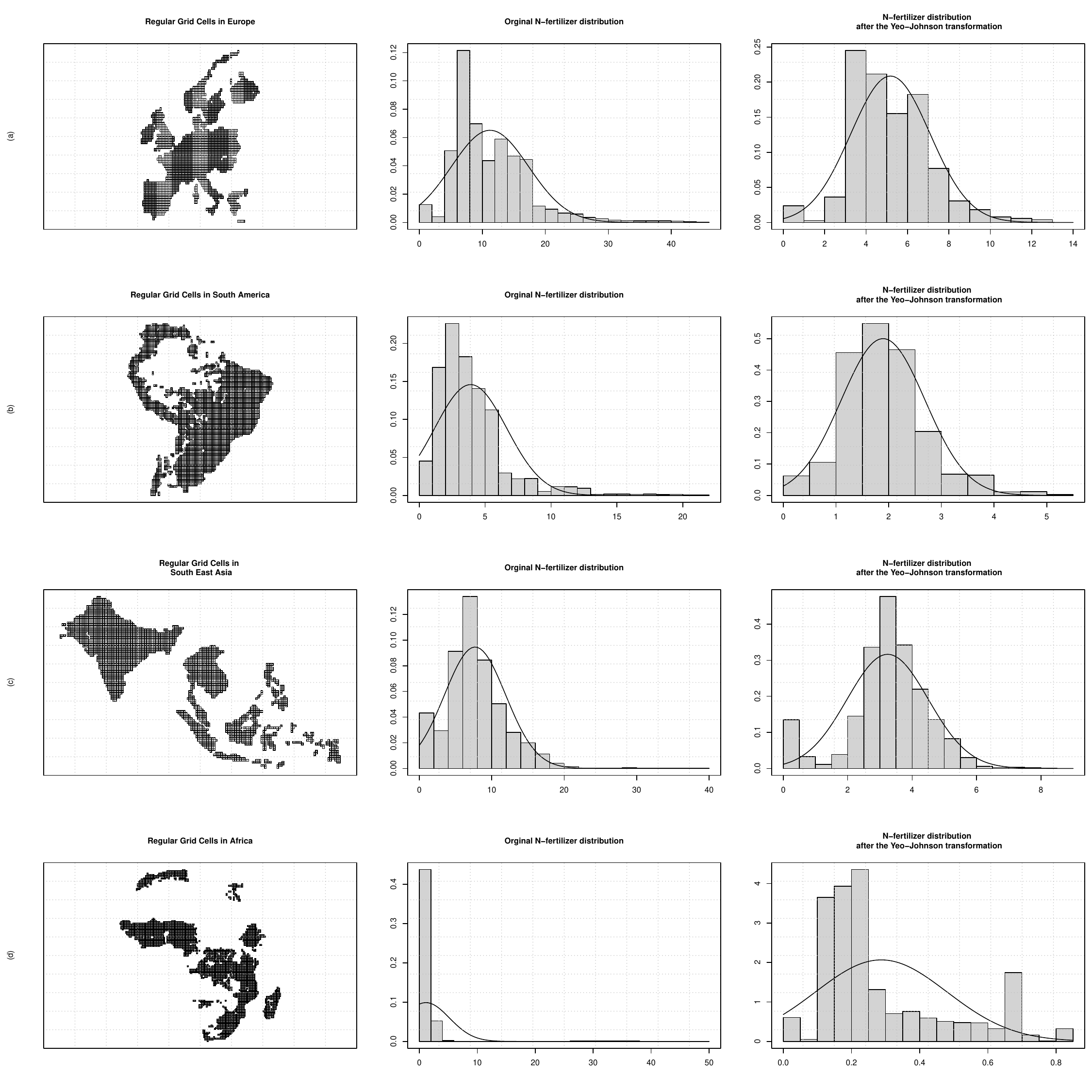}	 
\caption{Maps of (a) Europe (PAC zone), (b) South America, (c) South--East Asia and (d) Africa with their data distributions before and after YJ transformation (right-hand side). The values of $\lambda$ for the YJ transformations are equal to $\{0.55; 0.3; 0.4; -1.2\}$, respectively.}
\label{fig:fig1}	
\end{figure}

Even with a trimmed dataset in terms of considered years, the choice of a spatial dynamic panel data model, see Section \ref{sec:model}, renders too burdensome the computation of the model with all the data included. Furthermore, the strong climatic and socio--economic differences at world level calls for a subdivision of the data in more homogeneous zones. Finally, it is worth noting that comparing the results across developed and developing countries could be of particular interest for policy makers. Therefore, we selected four macro--regions over which to run the regressions separately: Europe (CAP zone), South America, South--East Asia and Africa. For the above four regressions, we consider balanced panel data sets, with the spatial dimension equal to $N=\{1928, 3508, 2703, 3493\}$ and the time dimension equal to $T=\{21,21,21,21\}$ years, respectively. Table \ref{tab:tab2} shows the considered countries included in each of these macro--regions together with the number of grid cells for each country. Figure \ref{fig:fig1}, instead, shows the spatial units (cells) taken into account for each macro--region and the approximated distributions of the N--fertilizer application with the Gaussian Kernel function. As we can observe, although the use of fertilization is by definition a truncated--at--zero Normal variable, justifying the use of a Tobit model\footnote{See the paper by \cite{xu2015maximum} for a spatial Tobit model in a cross-sectional setting.}, in all the four cases we can approximate quite well its distribution to a Gaussian, leading to the use of more complex and flexible spatio--temporal models already developed in the linear case, see Section \ref{sec:model}. To obtain more reliable Gaussian--type distributions, a Yeo--Johnson power transformation \citep{yeo2000new} of the dependent variables is also adopted and a comparison of the results is reported in Section \ref{sec:robust}.

Since the aim of the paper is to investigate the effect of weather conditions on the application of fertilizer, the explanatory variable of interest must be an indicator of their variation. We limit the analysis to dryness and wetness given the availability of reliable indexes measuring these conditions that further have the advantage to be computed at the same spatial level as our dependent variable. Different indexes with the ability to measure them are present in the literature and are potentially suitable for our analysis. Among them, we have the \href{https://crudata.uea.ac.uk/cru/data/drought/}{self-calibrated Palmer Severity Drought Index} (scPSDI) \citep{wells04}, the \href{https://iridl.ldeo.columbia.edu/maproom/Global/Precipitation/SPI.html}{Standardized Precipitation Index} (SPI) \citep{mckee93} and the \href{http://spei.csic.es/database.html}{Standardized Precipitation and Evapotranspiration Index} (SPEI) \citep{vicente10}. The scPSDI is a refinement of the Palmer Drought Severity Index (PDSI) \citep{palmer65}, one of the most widespread indexes used in drought related studies. The PDSI allows to measure both wetness and dryness -- positive values for the former and negative for the latter -- being based on the supply and demand concept of the water balance equation. This implies that it encompasses prior precipitation, moisture supply, run-off and evaporation demand at the surface level \citep{vicente10}. However, it has the drawbacks of being very sensitive to the temporal and spatial locations of calibration, of being scarcely comparable among different areas and of being rather subjective in relating drought conditions to the values of the index. A partial solution is provided by the scPDSI, that automatically calibrates the behaviour of the index at any location by replacing empirical constants in the index computation with dynamically calculated values \citep{wells04}. This process of self calibration increases the spatial comparability and reduces the subjectivity in evaluating extremes, but the index still retains the shortcomings of a strong autoregressive nature and a fixed time scale \citep{vicente10}. The SPI eliminates this last problem, but, on the other side, it only considers precipitation, disregarding other important characteristics such as temperature, evapotranspiration, wind speed and the capacity of soil to retain moisture. Finally, the SPEI, proposed more recently, tries to take into account these last elements and to combine the strength points of the previous indexes \citep{vicente10}. For these reasons, we decided to use this last index in our analysis.

The SPEI index is expressed as the number of standard deviations of weather conditions from the long term average, with negative values indicating dryness and positive values wetness. The yearly average of such index has been adopted in order to conform it with the time dimension of the dependent variable. In order to improve the intelligibility of the results, we have divided the SPEI index into two variables, dryness and wetness, with the first being equal to the SPEI index when it is lower than zero and assuming the value of zero otherwise. Dryness values have been subsequently multiplied by minus one in order to have only positive values. Wetness is simply defined as the complement of dryness. Figure \ref{fig:fig2} shows their distribution in the whole dataset excluding the zero values for readability.      

The countries' per capita Gross Domestic Product (GDP) is a fundamental control. In fact, the relation between the economic level of a country and the amount of fertilizer applied per unit of land is a well established fact in the dedicated literature. Furthermore, its interaction with the lagged term of dryness and wetness serves to test if the response of fertilization to past weather conditions changes according to level of income, as the literature on poverty traps suggests \citep{barrett07}. Since the purchase and application of fertilizer is likely to vary according to the types of grown crops and according to other local specificities, it may well be that it is the income of the past year rather than the one of the current year to affect the most this decision. For this reason, the time lag of per capita GDP is also included. This last may also evidence poverty traps more directly than the interaction of GDP with lagged weather conditions. Finally, in order to have a meaningful inter--countries comparability of the GDP, we adopted the per capita Power Purchasing Parity (PPP) GDP in constant 2011 dollars, provided by the \href{https://data.worldbank.org/indicator/ny.gdp.pcap.pp.kd}{World Bank}. 

We finally included other two relevant determinants to avoid a potential omitted variable bias problem, i.e. the price of agricultural output (PAO) and the price of fertilizer (PF). Lacking data at such a fine extension as our grid cells, we rely on country data with a generic price index for agricultural output used to broadly capture potential price effects. The limits of this index, obtained from the Food and Agriculture Organization database (\href{http://www.fao.org/faostat/en/#data}{Faostat}), are several since it basically considers an average of all crops prices, thus disregarding the specificities of each plot. It is, however, a second best option due to the territorial extension covered by our analysis that leaves no feasible alternatives. Table \ref{tab:tab1} reports the summary statistics of all the variables included in our model specifications, distinguishing them across macro--regions. Regarding PF, the introduction of this variable requires to considerably reduce the covered time span (all years before 2002 are dropped) and to drop some countries, due to the lack of data. We have therefore decided to include this variable in a separate analysis, while accounting also for potential spatial error dependences. There are also some caveats that must be remembered regarding this variable. First of all, our dependent variable is defined as the amount of nitrogen present in the soil, with nitrogen being one of the three macro--nutrients present in different fertilizers. Nitrogen, therefore, does not have a proper price, being this defined only for fertilizers. These lasts differ in the proportions of macro--nutrients they provide. In order to circumvent the problem of lack of price for nitrogen, we use as a proxy the price of Urea, a common fertilizer whose main component is nitrogen\footnote{Other common sources of nitrogen are Ammonia and Ammonium nitrate. The choice of Urea is simply due to a greater availability of data.}. The price of Urea, at country level, has been also collected from \href{http://www.fao.org/faostat/en/#data}{Faostat}. However, such database only provides information about import and export quantities and values together with the domestic used quantity. We use therefore the unit price of imported Urea -- obtained dividing the imported value by the imported quantity -- as a proxy for its domestic market price. It is finally worth noting that both these prices have required a procedure of interpolation in order to fill some missing values described in subsection \ref{sec:interpolation}.

%
\subsection{Data Interpolation}
\label{sec:interpolation}
We found several missing values over time in both the price of agricultural outputs (PAO) and the price of fertilizer (PF), especially for the latter one. To avoid the elimination of a large number of cells as well as to allow for the inclusion of these relevant exogenous regressors in our model specification, we opted to proceed by using some {\em ad hoc} data interpolation methods and filling the missing values. 

We first eliminated those cells belonging to those countries with more than the $30\%$ of missing values, i.e. the series are not large enough and they do not have sufficient information to produce accurate interpolations especially for those with higher volatility. For the PAO cases, only Africa lost some countries, i.e. Angola, Benin, Central African Republic, Chad, Congo, Gabon, Guinea--Bissau, Lesotho, Liberia, Mauritania, Sierra Leone, Sudan, Swaziland, Uganda, Zambia, Zimbabwe, whereas the PF series lost Lao PDR and Nepal in South East Asia and Benin, Central African Republic, Chad, Congo, Dem. Republic Congo, Eq. Guinea, Gabon, Guinea--Bissau, Lesotho, Liberia, Mauritania, Sierra Leone, Sudan and Swaziland in Africa. Data interpolation methods are then used according to the characteristic of the series and the number of missing values within the same series. In particular, we used trend interpolations with first--order polynomials, local interpolations with moving average and different orders $q$ (MA($q$)), and double or two--step interpolations that consist in filling some missing values in the first step and then using also the interpolated data to fill the others within the same series in the second step. 
\section{Model Specification}
\label{sec:model}
In this Section we provide details on the model specification to study the effect of extreme weather conditions on the use of fertilizers in Europe, South America, South--East Asia and Africa. More in general we try to specify the most appropriate and flexible econometric model, up to now, in order to model the N--fertilizer use and its main determinants. 

The adopted methodological strategy is based on the use of a {\em spatial dynamic panel data (SDPD)} model \citep{lee2010spatial, lee2014efficient}, or a {\em time--space dynamic} model \citep[p. 646]{anselin2008spatial}, to deal with both space and time dependence effects, i.e. spatial (cross--sectional) and serial correlations, as well as both space (individual) and time fixed effects. We also consider the possibility of including/excluding some variables depending on several theoretical and statistical issues like: (i) avoiding possible model overspecification and identification problems, (ii) avoiding inconsistency due to omitted variable biases, and (iii) avoiding inefficiency of irrelevant variables, especially those for which there is not an economic justification. 

The temporal lagged term, $\by_{n,t-1}$, is of particular interest in our context, since fertilizer utilization in each cell is likely to be affected by itself one year before. We then exclude a priori the possibility of using static specifications. We also consider fixed effects models rather than the random ones \citep{parent2012spatial, li2020spatial}: the former seem to be preferred due to computational and robustness reasons \citep{lee2010some}, especially when considering spatial panel specifications \citep[Sec. 3.4]{elhorst2014spatial}. Moreover, individual (spatial) fixed effects $\balpha_n$ (a specific form of {\em unobserved spatial heterogeneity}) could account for specific characteristics, like the soil type of the spatial units, for which we cannot obtain sufficient information. Time fixed effects $\xi_t\iota_n$ (a specific form of {\em unobserved time heterogeneity}) are instead able to capture all the phenomena that change over time and simultaneously affect the units in space, like e.g. an economic crisis. In fact, the omission of spatial and time fixed effects could bias the estimates in spatial and time series, respectively\footnote{It is worth noting that, one can use the Hausman test among static spatial panel data models \citep{mutl2011hausman} and SDPD models \citep{lee2020initial}, in order to discriminate between fixed and random specifications.}. 

The (contemporaneous) spatial lagged and the space--time lagged variables, i.e. $\sW_n \by_{n,t}$ and $\sW_n \by_{n,t-1}$, are also relevant determinants of fertilizer utilization worldwide. The source of the spatial dependence is to be found in a potential {\em social interaction} among farmers such as {\em peer effects}, especially in those macro--regions where there is a low level of know--how and the only source of knowledge is though mere communication among farmers about their own experiences. This type of interaction can in turn provide positive/negative spillover effects (e.g. of agricultural knowledge) over space and over time. Regardless, including spatial dependence in the dependent variables can avoid potential omitted variable bias problems and add relevant flexibility to the model specification. Finally, we exclude both (contemporaneous) spatially lagged and space--time lagged regressors, i.e. $\sW_n \bX_{n,t}$ and $\sW_n \bX_{n,t-1}$, since the regressors used in our empirical context are at least country--specific and therefore do not exhibit sufficiently variation at our disaggregation level of the data, leading to a redundance of irrelevant information rather than excluding a potential omitted variable bias problem. For the reason just explained, there is no an interesting economic meaning in the interpretation of spatially lagged GDP, dryness and wetness variables. Even when at least one of them is statistically significant, the reason could be more in the identification of a common factor that cannot be interpreted as a local spillover effect. Results related to robustness checks of our model are referred to Section \ref{sec:robust}.

We start by considering the following SDPD model with both individual and time fixed effects
\begin{equation}\label{eq:model}
\by_{n,t} = \rho \sW_n \by_{n,t}  + \phi \by_{n,t-1} + \gamma \sW_n \by_{n,t-1} + \bX_{n,t}\bbeta + \balpha_n + \xi_t\iota_n + \bvarepsilon_{n,t}, \;\; t=1,\dots,T
\end{equation}
where $\by_{n,t} = \left(y_{1,t}, y_{2,t}, \dots,  y_{i,t}, \dots, y_{n,t}\right)^\prime$ is a $n$--dimensional column vector of fertilizer utilization at time $t$, $\phi$ is the temporal autoregressive coefficient, $\sW_n$ is a time--invariant $n$--dimensional square matrix of spatial weights among pairs of random variables $\left(y_{i,t},y_{j,t}\right)$, for $i,j = 1,\dots,n$, with (contemporaneous) spatial autoregressive coefficient $\rho$ and space--time autoregressive coefficient $\gamma$, $\bX_{n,t} = \left(\x_{1,t}, \x_{2,t}, \dots, \x_{h,t}, \dots, \x_{k,t}\right)$ is an $n$ by $k$ matrix of non stochastic regressors including contemporaneous, time lagged\footnote{The time lagged gdp is included to control for a potential {\em poverty trap} problem or simply to the fact that farmers apply the fertilizer bought the year before.} and squared gross domestic product, i.e. $GDP_t$, $GDP_{t-1}$, $GDP^2_t$, contemporaneous and time lagged truncated (at zero) normal variables for dryness and wetness, i.e. $DRY_t$, $WET_t$, $DRY_{t-1}$, $WET_{t-1}$, contemporaneous and time lagged agricultural output\footnote{The contemporaneous price of agricultural output is included to control for a potential omitted variable problem.}, i.e. $PAO_t$ and $PAO_{t-1}$, and interaction terms like $GDP_t\times DRY_{t-1}$ and $GDP_t\times WET_{t-1}$ with $\bbeta$ the vector of coefficients, $\balpha_n$ is an $n$--dimensional column vector of spatial (individual) fixed effects, $\xi_t\iota_n$ is an $n$--dimensional column vector of time fixed effects with scalar coefficient $\xi_t$ and column vector of ones $\iota_n$, and $\bvarepsilon_{n,t} = \left(\varepsilon_{1,t}, \varepsilon_{2,t}, \dots, \varepsilon_{i,t}, \dots, \varepsilon_{n,t}\right)^\prime$ is an $n$--dimensional column vector of innovations at time $t$ with $\varepsilon_{i,t}$ independent and identically distributed (i.i.d.) across $i$ and $t$ with zero mean and finite variance $\sigma^2$.

The time--invariant spatial weighting matrix $\sW_n = \{w_{ij}\}$ is a row--stochastic matrix such that
$$\begin{cases} w_{ij} = \frac{1}{\sum_j w_{ij}} \quad iff \quad y_j \in \mathcal{N}_k \\ w_{ij} = 0 \qquad \quad otherwise \end{cases}$$
where $\mathcal{N}_k$ is the set of nearest random variables $y_j$ to $y_i$ defined by $k$. In our case we set $k=4$, which can be quite similar to the queen contiguity scheme for regular square lattice grids and several units with no more than 5 neighbours on average\footnote{See Section \ref{sec:robust} for robustness checks on the use of other weighting schemes.}. The reason why we use the above $k$--nearest neighbour approach is based on the fact that it guarantees the equivalence between the two spatial dynamic models defined before and after row--normalization of the weights. Moreover, $k=4$ is a reasonable choice looking at the spatial distribution of the cells and assuming that the direct strategic interaction effects among farmers are not highly relevant at greater distances. The set of nearest neighbours for each unit in space is defined through the Euclidean distances between pairs of centroids of the grid cells.
\subsection{Stability and First--differencing}
\label{sec:stability}
To ensure stable spatio--temporal processes the condition $\rho + \phi + \gamma < 1$ must be satisfied. During a preliminary estimation procedure of equation \eqref{eq:model}, we found the above condition to be numerically satisfied for the 4 macro--regions. However, all the four sums are very close to the unit root, leading to potential inconsistency and numerical instability of the estimates. We conducted (two--sided) Wald tests on the null hypothesis that $\rho + \phi + \gamma = 1$ ({\em spatially cointegrated} processes) by using the statistic
$$W = \left(\rho + \phi + \gamma - 1\right)\left(1 \; 1 \; 1 \; \underline{0}^\prime\right)\Sigma_{QMLE}\left(1 \; 1 \; 1 \; \underline{0}^\prime\right)^\prime\left(\rho + \phi + \gamma - 1\right) \approx \chi_1$$
where $\Sigma_{QMLE}$ is the $\left(k + 4\right)$ by $\left(k + 4\right)$ covariance matrix of the {\em bias--corrected} QMLE estimator \citep{lee2010spatial,yu2008quasi}\footnote{The asymptotic bias--correction form is needed due to the joint estimation of both the individual and time fixed effects with the other parameters of interest in SDPD models. Alternatively, the GMM approach \citep{lee2014efficient} can be used, even if \cite{lee2020initial} have recently recommended the use of the QMLE for short SDPD models.}. In all the four cases we rejected the null hypothesis of spatial cointegration, but even so we opted to take the model in first--difference to avoid numerical instability\footnote{Indeed, especially in the European and South--East Asia area we found values very close to the unit root using R, i.e. 0.973 and 0.958, respectively, and even more using Stata.}. 

To remove inconsistency of the estimator in both spatial cointegration and spatial explosive cases, preserving also the amount of available observations, a {\em spatial first--differencing} approach has been recently proposed, see \cite{lee2010some} and \cite{yu2012estimation}, losing some degrees of freedom. Alternatively, one can re--estimate all the models after {\em time first--differencing}, losing one year of observations. The time--differencing is also useful to eliminate the individual fixed effects and, therefore, the inconsistency due to their correlation with $y_{t-1}$ \citep[page 100]{elhorst2014spatial}, or a demeaned version can be also used. 

By re--writing the model in equation \eqref{eq:model} with the explicit inclusion of all the regressors, and suppressing $n$ for notational convenience, we obtain the following two specifications form time first--differencing and spatial first--differencing approaches, respectively. Defining $\Delta = \left(\bI - \sL\right)$ the time first--differencing operator such that $\Delta\bv_t = \bv_t - \bv_{t-1}$ with $\bv_t$ a variable vector at time $t$, we obtain
\begin{align}\label{eq:timediff}
\Delta\by_t &= \rho \sW \Delta\by_t  + \phi \Delta\by_{t-1} + \gamma \sW \Delta\by_{t-1} + \beta_1\Delta\x_{1,t} + \beta_2\Delta\x_{1,t}^2 + \beta_3\Delta\x_{2,t} + \beta_4\Delta\x_{3,t} + \beta_5\Delta\x_{4,t}+ \beta_6\Delta\x_{1,t-1} + \cr
& \qquad + \beta_7\Delta\x_{2,t-1} + \beta_8\Delta\x_{3,t-1} + \beta_9\Delta\x_{4,t-1} + \beta_{10}\Delta\x_{1,t}\x_{2,t-1} + \beta_{11}\Delta\x_{1,t}\x_{3,t-1} + \Delta\xi_t\iota + \Delta\bvarepsilon_t
\end{align}
where $\left(\x_{1,t},\x_{2,t},\x_{3,t},\x_{4,t}\right)$ are referred to $GDP_t$, $DRY_t$, $WET_t$ and $PAO_t$, respectively, and the individual (spatial) fixed effects $\balpha$ are suppressed since they are time--invariant, i.e. $\Delta\balpha = \b0$. In the same way, defining $\Gamma = \left(\bI - \sW\right)$ the spatial first--differencing operator, we obtain 
\begin{align}\label{eq:spatialdiff}
\Gamma\by_t &= \rho \sW \Gamma\by_t  + \phi \Gamma\by_{t-1} + \gamma \sW \Gamma\by_{t-1} + \beta_1\Gamma\x_{1,t} + \beta_2\Gamma\x_{1,t}^2 + \beta_3\Gamma\x_{2,t} + \beta_4\Gamma\x_{3,t} + \beta_5\Gamma\x_{4,t}+ \beta_6\Gamma\x_{1,t-1} + \cr
& \qquad + \beta_7\Gamma\x_{2,t-1} + \beta_8\Gamma\x_{3,t-1} + \beta_9\Gamma\x_{4,t-1} + \beta_{10}\Gamma\x_{1,t}\x_{2,t-1} + \beta_{11}\Gamma\x_{1,t}\x_{3,t-1} + \Gamma\balpha + \Gamma\bvarepsilon_t
\end{align}
where $\left(\x_{1,t},\x_{2,t},\x_{3,t},\x_{4,t}\right)$ are defined as before and the time fixed effects $\xi_t\iota$ are suppressed since they are spatial--invariant, i.e. $\Gamma\xi_t\iota = \b0$. The model in equation \eqref{eq:spatialdiff} exhibit heteroskedasticity, i.e. the covariance matrix is $\Gamma\Gamma^\prime$ which is singular. The used likelihood \citep[eq. 5.19, page 96]{lee2011estimation}, therefore, involves the generalized inverse of $\Gamma\Gamma^\prime$, which is equal to the matrix multiplication of the appropriate eigenvector and eigenvalues matrices obtained form its spectral decomposition, and after the elimination of those eigenvectors whose eigenvalues are zero.

Finally, for both the models in equations \eqref{eq:timediff} and \eqref{eq:spatialdiff}, a {\em within (or demeaned) transformation} is used to eliminate the transformed time fixed effects and the transformed individual fixed effects, respectively. Note that within transformations induce (spurious) cross--sectional and serial error correlations, from cross--sectional and time demeaning, respectively \citep[eq. 10.52, page 270]{wooldridge2010econometric}. Since the error correlations for the above two models will be equal to $-\frac{1}{T-1}$ and $-\frac{1}{N-1}$ which disappear as $T\to\infty$ and $N\to\infty$, respectively, we suppose they are negligible with the use of our big panel data sets. The estimation results after spatial first--differencing are reported in a table in the Supplementary material. Careful attention should be paid on these results. For instance, it is quite suspected the value of $\rho$ in Europe close to the boundary. Moreover, although the spatial first--differencing approach can be seen as a general approach, in our case we found the sums $\rho + \phi + \gamma = \{0.97318, 0.93883, 0.95820, 0.92282\}$ and the Wald statistic values $\{50.6686, 264.1347, 167.2486, 400.2887\}$ rejecting all the null hypotheses of spatial cointegration in Europe, South America, South East Asia and Africa, respectively. In this paper, we finally decide to report all the results in time first--differencing, leaving a more accurate statistical comparison of the two first--differencing approches to further researches.
\subsection{Controlling for Spatial Error Correlations}
\label{sec:sarar}
Although the model in equation \eqref{eq:model}, or equivalently \eqref{eq:timediff}, is considered quite general in its form, in this paper we also allow for the possibility of the error terms to be spatially correlated. Several statistical hypothesis testing that check for the presence of potential spatial error autocorrelations in panel data specifications have been proposed, see e.g. \cite{baltagi2003testing}, \cite{millo2017simple}. However, none of them consider a spatial dynamic specification of the panel data model under the null hypothesis, excluding the usability of the statistics for model in equation \eqref{eq:model}. An alternative is to consider the evolution of the Moran's Index (I) and hypothesis testing \citep{moran1950notes} over time. The Moran's statistic, $-1 < I <1$, can be written as  
\begin{align}\label{eq:moran}
I_t = \frac{n}{\sum_i \sum_j w_{2,ij}}\frac{\sum_i \sum_j w_{2,ij}\left(\varepsilon_{i,t} - \bar{\varepsilon}_t\right)\left(\varepsilon_{j,t} - \bar{\varepsilon}_t\right)}{\sum_i \left(\varepsilon_{i,t} - \bar{\varepsilon}_t\right)^2} \quad \forall t
\end{align}
whose normalized version has a Normal distribution in each year, and where $w_{2,ij}$ are the elements of a time--invariant weight matrix. Once the presence of the spatial error autocorrelation has been detected, the model in equation \eqref{eq:model} can be extended as follows
\begin{align}\label{eq:sarar}
\by_{n,t} = \rho \sW_{1,n} \by_{n,t}  + \phi \by_{n,t-1} + \gamma \sW_n \by_{n,t-1} + \bX_{n,t}\bbeta + \balpha_n + \xi_t\iota_n + \bu_{n,t}, \quad \bu_{n,t} = \lambda \sW_{2,n}\bu_{n,t} + \bvarepsilon_{n,t}
\end{align}
where $\sW_{1,n} = \sW_n$ in equations \eqref{eq:model} and \eqref{eq:timediff} whose elements are defined through the $k$--nn approach with $k=4$, while $\sW_{2,n}$ is the time--invariant weighting matrix used for the multivariate error process with $\lambda$ its autoregressive coefficient. The elements of $\sW_{2,n}$ are also defined through the $k$--nn approach but $k=18$, to avoid identification problems that generally occur when $\sW_{1,n} = \sW_{2,n}$. Also in this case, the choice of $k=18$ for the error process seems to be reasonable looking at the spatial distribution of the cells. Indeed, for this type of process we assume the impact of a shock in one site can directly propagate over a greater number of neighbours, like e.g. a contamination of the soil. Discussions on the Moran's I, his hypothesis testing and results on SDPD models with error correlations are referred to subsection \ref{sec:error_priceur}.
\subsection{Time--varying Marginal Effects}
\label{sec:meffects}
According to the spatial econometric literature, proper marginal effects that take spatial effects into account must be defined in different ways depending on the specified model. When considering SDPD models, the definition of the total, direct and indirect (spillover effects) impacts are also different with respect to the short--term and long--term periods, see \cite{debarsy2012interpreting} and \citet[Sec. 4.6]{elhorst2014spatial}. Considering the model in equation \eqref{eq:timediff} with time first--differencing, the marginal effects can be defined, as for levels, in the following way
\begin{align}\label{eq:marginal_short}
\frac{\partial \mathbb{E}\left(\by_t\right)}{\partial \x_{1,t}}\mid_t &= \left(\bI - \rho\sW\right)^{-1}\left[\beta_1\bI+2\beta_2\diag{(\x_{1,t})}\bI+\beta_{10}\diag{(\x_{2,t-1})}\bI+\beta_{11}\diag{(\x_{3,t-1})}\bI\right] \nonumber \\
\frac{\partial \mathbb{E}\left(\by_t\right)}{\partial \x_{2,t}}\mid_t &= \left(\bI - \rho\sW\right)^{-1}\left[\beta_3\right] \nonumber \\
\frac{\partial \mathbb{E}\left(\by_t\right)}{\partial \x_{3,t}}\mid_t &= \left(\bI - \rho\sW\right)^{-1}\left[\beta_4\right] \nonumber \\
\frac{\partial \mathbb{E}\left(\by_t\right)}{\partial \x_{4,t}}\mid_t &= \left(\bI - \rho\sW\right)^{-1}\left[\beta_5\right] 
\end{align}
for the short--term impacts, and 
\begin{align}\label{eq:marginal_long}
\frac{\partial \mathbb{E}\left(\by_t\right)}{\partial \x_{1,t}} &= \left[\left(1-\phi\right)\bI - \left(\rho+\gamma\right)\sW\right]^{-1}\left[\beta_1\bI+2\beta_2\diag{(\x_{1,t})}\bI+\beta_{10}\diag{(\x_{2,t-1})}\bI+\beta_{11}\diag{(\x_{3,t-1})}\bI\right] \nonumber \\
\frac{\partial \mathbb{E}\left(\by_t\right)}{\partial \x_{2,t}} &= \left[\left(1-\phi\right)\bI - \left(\rho+\gamma\right)\sW\right]^{-1}\left[\beta_3\right] \nonumber \\
\frac{\partial \mathbb{E}\left(\by_t\right)}{\partial \x_{3,t}} &= \left[\left(1-\phi\right)\bI - \left(\rho+\gamma\right)\sW\right]^{-1}\left[\beta_4\right] \nonumber \\
\frac{\partial \mathbb{E}\left(\by_t\right)}{\partial \x_{4,t}} &= \left[\left(1-\phi\right)\bI - \left(\rho+\gamma\right)\sW\right]^{-1}\left[\beta_5\right]
\end{align}
for the long--term impacts. The averages of the diagonal elements of the matrices in equations \eqref{eq:marginal_short} and \eqref{eq:marginal_long} are the direct impacts, whereas the off--diagonal averages of the matrices in equations \eqref{eq:marginal_short} and \eqref{eq:marginal_long} are the indirect impacts. Note that, in our case, the short--term and long--term effects with respect to $\x_1$ ($GDP$) are time--varying, and their evolution is shown in Figure \ref{fig:tv_me}.

In addition, we consider specific marginal effects obtained from the following error correction model (ECM) representation of equation \eqref{eq:timediff} \citep{yu2012estimation}
\begin{align}\label{eq:ecm}
\Delta^2 \by_t &= \rho \sW \Delta^2 \by_t + \left(\phi - 1\right)\Delta\by_{t-1} + \left(\rho + \gamma\right)\sW\Delta\by_{t-1} + \beta_1\Delta\x_{1,t} + \beta_2\Delta\x_{1,t}^2 + \beta_3\Delta\x_{2,t} + \beta_4\Delta\x_{3,t} + \beta_5\Delta\x_{4,t} + \cr
& \qquad + \beta_6\Delta\x_{1,t-1} + \beta_7\Delta\x_{2,t-1} + \beta_8\Delta\x_{3,t-1} + \beta_9\Delta\x_{4,t-1} + \beta_{10}\Delta\x_{1,t}\x_{2,t-1} + \beta_{11}\Delta\x_{1,t}\x_{3,t-1} + \Delta\xi_t\iota + \Delta\bvarepsilon_t
\end{align}
%
from which we can easily calculate the following marginal impacts of interest 
\begin{equation}\label{eq:marginal_ecm}
\frac{\partial \mathbb{E}\left(\Delta^2 \by_t\right)}{\partial \Delta\by_{t-1}} = \left(\bI - \rho\sW\right)^{-1}\left[\left(\phi - 1\right)\bI + \left(\rho + \gamma\right)\sW\right]
\end{equation}
which are the effects estimates of convergence, and 
\begin{align}\label{eq:marginal_ecm2}
\frac{\partial \mathbb{E}\left(\Delta^2 \by_{t}\right)}{\partial \Delta\x_{2,t-1}} &= \beta_7\bI + \beta_{10}\diag{(\x_{1,t})}\bI \nonumber \\
\frac{\partial \mathbb{E}\left(\Delta^2 \by_{t}\right)}{\partial \Delta\x_{3,t-1}} &= \beta_8\bI + \beta_{11}\diag{(\x_{1,t})}\bI 
\end{align}
which are specific effects related to the variables dryness and wetness at time $t-1$. Since these last two effects are also time--varying, their evolution is shown in Figure \ref{fig:tv_ecm_me}. Finally, since spatial marginal effects can also be referred to each statistical unit in space to reveal a source of heterogeneity, in Figures \ref{fig:fig4}, \ref{fig:fig5}, \ref{fig:fig6}, and \ref{fig:fig7} we report maps on these effects. See \ref{sec:local_me} for details on their calculation.
\section{Results and Discussion}
\label{sec:results}
In this Section we report and discuss our main estimation results and the potential policy implications derived from them. For the estimation results, we used the function {\em spml} in the R package \texttt{splm} \citep{millo2012splm}. Alternatively, Stata command \texttt{xsmle} \citep{belotti2017spatial} and Matlab codes at \texttt{https://spatial-panels.com/software/} \citep{elhorst2013impact} can be used. All the other calculations are instead implemented by ourselves in R. For the estimation of the spatial first--differencing model in equation \eqref{eq:spatialdiff} we used the {\em bobyqa} optimization function in the R package \texttt{minqa} \citep{nash2011unifying}. As already mentioned in subsection \ref{sec:stability} a {\em within (or demeaned) transformation} for all the considered models has been adopted. In Table \ref{tab:tab3}, we show the coefficients estimates of model in equation \eqref{eq:timediff} for all the four regressions\footnote{Estimation results of the ECM in eq. \eqref{eq:ecm} are available upon request. In this paper we focus the attention on model in eq. \eqref{eq:timediff}, although we provide marginal effects also for the model in eq. \eqref{eq:ecm}.}, whereas the results with the dependent variable having been YJ transformed are available in the Supplementary material of this paper.

First of all, an important role is played by both the spatial, temporal and spatio--temporal variables in our model specification. As expected, the amount of fertilizer in a previous year is negatively correlated with the level of N--fertilization applied in the current year ($\phi$). The more nutrients are applied in a given year, the lower will be the need one year later. This reasonable result applies both to developed and developing regions. The (contemporaneous) spatial autoregressive coefficient ($\rho$) is strongly significant and positive in all the macro--regions, implying that the spatial process is not inhibitory and that the choice of the fertilizer quantity is rather equivalent among close units of lands, as largely expected. This may be due to both the similarity of agro--climatic conditions of close units of lands, thus implying the cultivation of similar crops and the adoption of analogous agronomic practices, and potential knowledge spillover effects among farmers. The spatio--temporal coefficient ($\gamma$) is also positive and significant in all the regressions but with a lower magnitude than the one of the spatial autoregressive coefficient. The positiveness of both coefficients strengthens the hypothesis of knowledge spillovers among farmers located in neighbour areas. Exchange of information through direct contacts or through farmers associations are examples of how these knowledge spillovers could take place. If the contemporaneous effect may be due to similar environmental conditions, the temporal lagged one seems in fact to indicate the persistence of an imitation effect over time. Therefore, knowledge spillovers are clearly stronger at present time, but they also persist for, at least, one year.

Looking at the effect of current weather conditions, the coefficients are significant in Europe and in South America. In the first case, only wetness is significant and positive, whereas in South America it is significant for both deviations, with the coefficient for wetness being negative -- opposite to Europe -- and the one for dryness positive. The lower efficacy of certain crops in absorbing N--fertilizer during wet conditions, described in \citet{vlek86}, could be the reason explaining the positive coefficient in Europe, whereas the greater need of N--fertilization under drought conditions evidenced by \citet{purcell04} provides a reason for the positive value of the dryness coefficient in South America. The negative coefficient for wetness in this last region, instead, may be due to the diffusion of crops or crop varieties less negatively affected in their capacity to absorb N--fertilizer under wet conditions combined with a substitution effect. In particular, in presence of wet conditions that favour the growth of crops, farmers would tend to rely on them and saving inputs. It is questionable, however, that this is a sounding agronomic practice. When considering their lagged terms, the coefficients are still significant only in Europe and South America\footnote{Actually, in Africa the lagged term of wetness is significant when considering the YJ transformation, see Tables in the Supplementary material.}. The negative effect of past dryness is consistent with the lower need of fertilization after a drought period due to the availability in the soil of the nutrients left from the past season (see footnote \ref{note:note1}). This is therefore an optimal strategy, apparently followed only in the most industrialized region. The increase of fertilization in South America as a consequence of past excesses of wetness is also reasonable and justified from an agronomic point of view since wet conditions may favour the run--off of nutrients from the soil. The opposite coefficient of this variable in Africa – significant only under the  YJ transformation – may instead be the result of wrong farming practices, see Tables in the Supplementary material. The almost total lack of significance of the climatic variables in Africa and South East Asia deserves attention. This could be due to a simple technical reason, e.g. a lower quality of data. However, it could also be due to a lower ability of local farmers to adjust the input quantities to both present and past climatic variations. This may be due to a lack of proper training or proper means to analyse the soil and calls for an improvement in extension services. 

Regarding the other relevant control variables, the comments are as follows. The role played by per capita GDP varies considerably between the examined macro--regions. Its effect at time $t$ follows a concave downward parabola in Europe. This is consistent with the presence of an environmental Kuznets curve for fertilization, hypothesis already confirmed by \citet{li16} for China. Moreover, the lagged term is negative. If we consider the time of purchase of fertilizers as being dependent from the grown crop type and its related optimal period of fertilization, rather than from the immediate availability of economic means, as it seems plausible in an industrialized region, it is legitimate to expect a similar role of GDP at time $t$ and $t-1$. This seems to be true in our case, confirming a tendency of wealthier countries to adopt more stringent environmental policies, thus reducing the level of fertilization. On the contrary, Africa has not a significant contemporaneous per capita GDP coefficient, but a positive and significant lagged GDP. The positive role of a country's economic attainment on fertilization in this region is largely expected, whereas the significance of the lagged GDP deserves more attention. A plausible explanation is that strong financial constraints suffered by farmers oblige them to purchase fertilizers when they have cash at disposal, and this generally coincides with the sale of farmers' agricultural products.
Consequently, it is the last year economic level to shape the fertilization pattern of the current year since the sale of agricultural output happens at the end of the growing season. The positive coefficient of lagged GDP can also be interpreted as a {\em poverty trap problem}, since it originates from economic vulnerability and leads to underinvestment in productive assets after a shock. A similar conclusion can be drawn for South--East Asia. Here, The GDP at time $t$ is instead significant and negative, whereas the squared term is not significant. This last result may appear odd for this region, since invoking the role of environmental policies as justification of the negative relation is inconsistent with the positive role of the lagged term and, more generally, with the economic level of this region. Looking at the coefficient magnitudes, the contemporaneous effect is smaller than the lagged term. A plausible explanation is that several farmers are actually engaged in various activities, with off--farming ones being a better source of income. A lower level of per capita GDP is linked with lower possibilities for off--farming opportunities, thus increasing the time and the efforts dedicated to farming and consequently explaining the inverse relation between GDP and fertilization. The most puzzling result is the one of South America, where per capita GDP has a positive and significant coefficient for the level of the contemporaneous variable and negative for the one of the lagged term. 
Focusing on the contemporaneous effect, the positive -- linear or convex -- relation between fertilization and per capita GDP seems to resemble the one of a developing area laying in the increasing region of the Kuznets curve. However, the negative lagged term, contrasting the contemporaneous effect, is of difficult interpretation, thus deserving the attention of further studies.

Finally, the coefficient of the price of the agricultural output (PAO) at time $t$ is positive and significant in all the macro--regions, except for South--East Asia. If farmers are able to forecast an increase in the value of their output, it is then rational to increase the level of inputs to maximize production, thus justifying the positive relation between fertilization and PAO. The lagged term of this variable is instead significant and negative in all the regressions, except the one in Europe. Theoretically, a positive coefficient seems to be more reasonable for the lagged term too, especially for the developing countries. Indeed, a higher price of the agricultural output should imply higher profits for farmers, thus allowing them to have more resources to purchase inputs. However, since agricultural output includes all cultivated crops and several of them are used as fodder for livestocks, which in turn provide manure (a close substitute for chemical fertilizers), it might be that an increase in the agricultural output price leads to an increase in manure price and, consequently, in chemical fertilizer price too. In this specific case, the negative coefficient could then be a consequence of an omitted variable problem. In subsection \ref{sec:error_priceur}, where we will discuss the results of the regressions including the price of Urea, we will see that this interpretation seems highly likely. 

\subsection{Discussion on the Marginal Effects}
From an economic point of view, and for ad--hoc policy interventions, it is of main interest to take a look at the marginal effects. As already explained in Section \ref{sec:model}, marginal effects for SDPD models can be distinguished according to direct and indirect (due to spillovers) effects and short--term and long--term effects. For our model specifications in eq. \eqref{eq:timediff} and \eqref{eq:ecm}, we also provide a distinction between time--invariant marginal effects, see Table \ref{tab:tab5}, and time--varying marginal effects, see Figures \ref{fig:tv_me} and \ref{fig:tv_ecm_me}, due to the non--linearity in $GDP_t$ and the presence of additional interaction terms, respectively. 

Let us consider first the {\em time--varying short--term and long--term total marginal effects} with respect to per--capita GDP at time $t$ from model in equation \eqref{eq:timediff}, see also equations \eqref{eq:marginal_short} and \eqref{eq:marginal_long}, shown in Figure \ref{fig:tv_me}. First of all, there is a small difference between the short and long terms effects, such that it is rather useless to comment them separately. The temporal mean values for Europe, South America, South--East Asia and Africa, respectively, are around $\{0.06825; 0.07524; -0.04697; 0.00374\}$ for short--term direct effects, and $\{0.06605; 0.06851; -0.03817; 0.00333\}$ for long--term direct effects, while $\{0.00013; 0.00005; -0.00006; 0.000002\}$ and $\{0.00012; 0.00005; -0.00004; 0.000001\}$ for short--term indirect effects and long--term indirect effects, respectively. The magnitude of the indirect effects is so trivial in all the regions compared to the one of the direct effects, that it is again advisable to overlook them. This is somehow not surprising, however, since the variable $GDP$ do not vary at a unit--level but rather at a country--level, so that the indirect impacts due to neighbouring cells could be mitigated by the absence of variability in per--capita $GDP$ across units within the same country. The time--varying total marginal effects are always positive in all the macro--regions, with the exception of South East Asia where they are always negative\footnote{Obviously, the sign of the marginal effects directly depends on the combination of signs of the estimated coefficients that directly enters in the calculation of the marginal effects.}, and the absolute magnitude in Africa seems not to be very high compared with all the other macro--regions.

Figure \ref{fig:tv_ecm_me} reports the time--varying marginal effects with respect to the weather variables at time $t-1$ from the model in equation \eqref{eq:ecm}, see also equation \eqref{eq:marginal_ecm2}. The evolutions of both dryness and wetness are quite flat in all the considered macro--regions, but some differences can be found. In Europe and South America, the evolution of dryness is always negative with temporal mean values equal to $\{-0.04689;-0.00233\}$, respectively. The evolution of wetness is instead always positive with temporal mean values equal to $\{0.02149;0.02900\}$, respectively. These two area seem to reveal a similar evolution, although with different magnitudes and a slightly higher volatility in South America. Interesting are the lowest and highest peaks reached in 2009, just after the financial crisis started in 2007, where probably N--fertilization becomes much more sensitive to variations in per--capita GDP. South--East Asia shows a similar evolution for dryness and wetness, whose mean values are $\{-0.00299; -0.00240\}$, respectively. It is interesting to note that its highest negative peak is reached in 1998 during the financial crises of the so called Asian tigers. 
In Africa, instead, the evolution of dryness is flat compared to the one of wetness, with temporal mean values equal to $\{-0.00266; -0.00221\}$, respectively. 

Regarding the {\em time--invariant marginal effects} in Table \ref{tab:tab5}, we note that only in Europe the direct, indirect and total marginal effects are positive both in the short--term and in the long--term with respect to all the variables, whereas a variation of signs can be found in the other macro--areas. Therefore, in Europe, the higher are dryness, wetness and the price of agricultural outputs, the greater is the N--fertilization in both the cell itself and in the neighbour cells, whereas this is true in South America only for dryness and the price of agricultural outputs, in South East Asia only for wetness and in Africa only for the price. Looking at the estimates of convergence from the ECM, we can observe that both the strength of convergence of the cell itself and the strength of convergence of the other cells are negative in all the macro--areas, with a higher absolute value in South East Asia. Therefore, the higher is the use of N--fertilization the year before, the lower is the amount of N--fertilizer used a year later, corroborating the results of the coefficient of the SDPD model.  
Finally, in Figures \ref{fig:fig4}, \ref{fig:fig5}, \ref{fig:fig6}, and \ref{fig:fig7}, we show the spatial heterogeneity in terms of the time--invariant short--term marginal effects in equation \eqref{eq:marginal_short} with respect to the weather variables at time $t$. First of all, the range and the sign of the values do not necessarily correspond to the mean values in Table \ref{tab:tab5}, since in this case we only consider the indirect impacts as row--sums of the matrix in \ref{sec:local_me}. The spatial patterns seem to show a higher indirect impacts for both dryness and wetness in the majority of the grid cells, especially inside the core of all the macro--areas. This could mean that weather conditions, both in normal and in extreme cases, have a greater impact in those areas.
\subsection{Discussions on Spatial Error Correlations and Fertilizer Prices}
\label{sec:error_priceur}
In this Section we propose to re--estimate the model in equation \eqref{eq:timediff}, by controlling for spatial error autocorrelations to improve estimation efficiency and by including fertilizer (Urea) prices, since their omission could bias the estimates. Additionally, we also consider the impact coming from the same price at time $t-1$. This new model specification is defined in equation \eqref{eq:sarar}. The analysis is restricted to 10 years from 2004 to 2013, since no data are available for the years before 2002 and after having dropped 2002 and 2003 for time--differencing and the inclusion of the temporal lag. 

\begin{figure}[ht]	
	\centering
	\includegraphics[width=0.90\textwidth,height=0.80\textwidth]{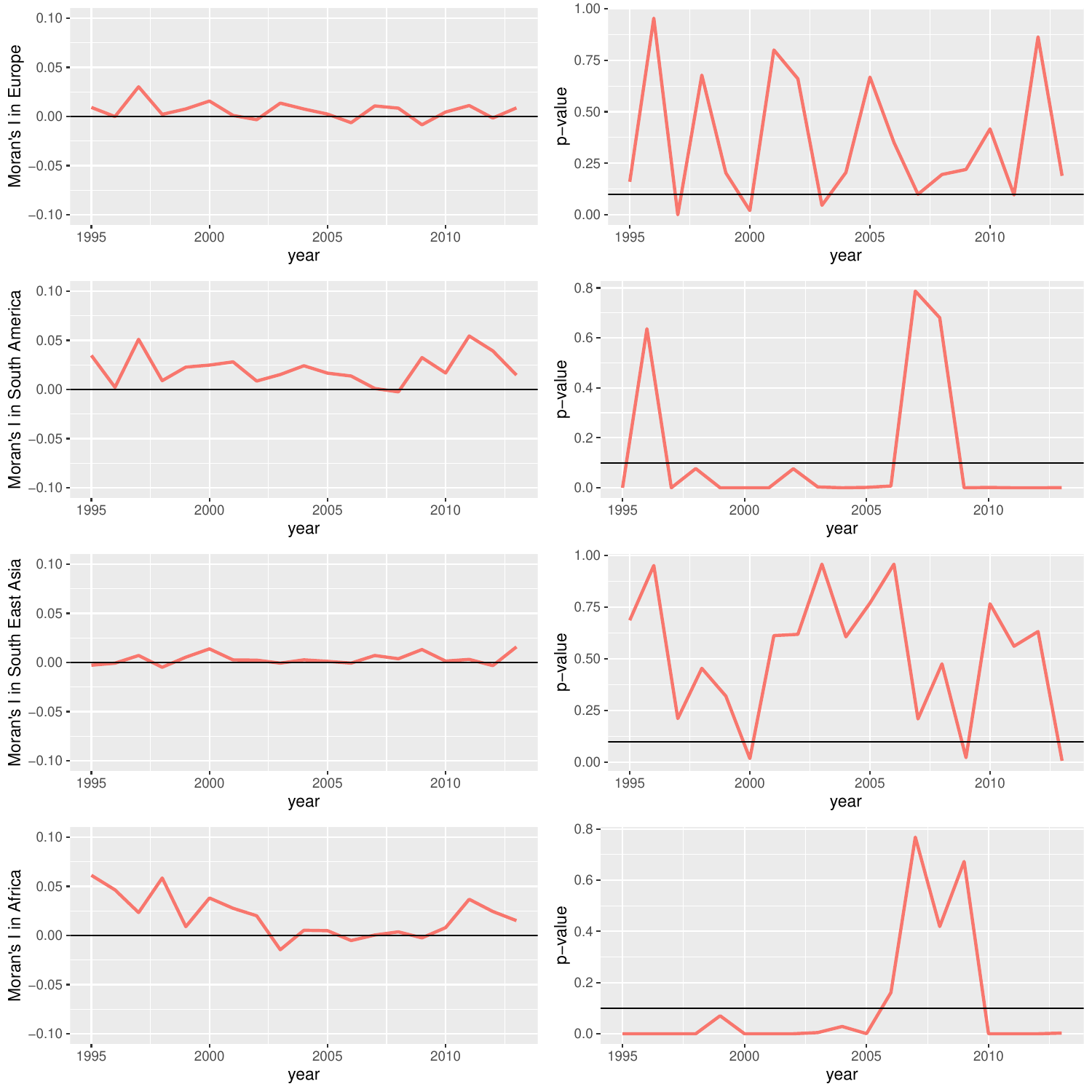}
\caption{Moran's I (left) and p-value of the (two--sided) Moran's Test (right) for the four macro--regions to detect the additional presence of spatial error autocorrelation in each year. On the right, red dotted lines represent $10\%$ significant level of the two--sided test.}
\label{fig:moran}	
\end{figure}

\begin{figure}[ht]	
	\centering
	\includegraphics[width=0.4\textwidth,height=150pt]{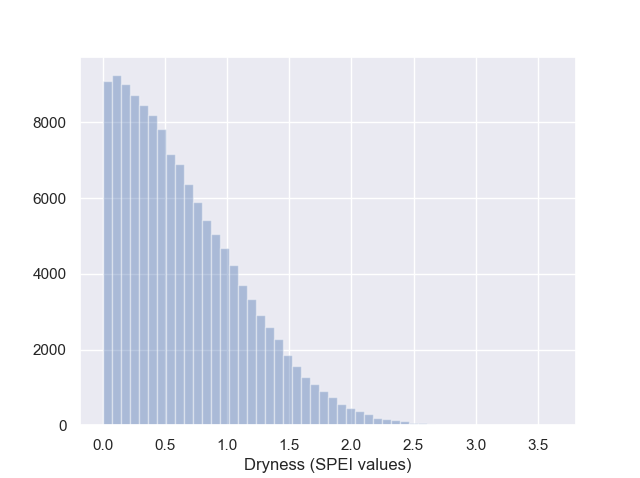}
	\includegraphics[width=0.4\textwidth,height=150pt]{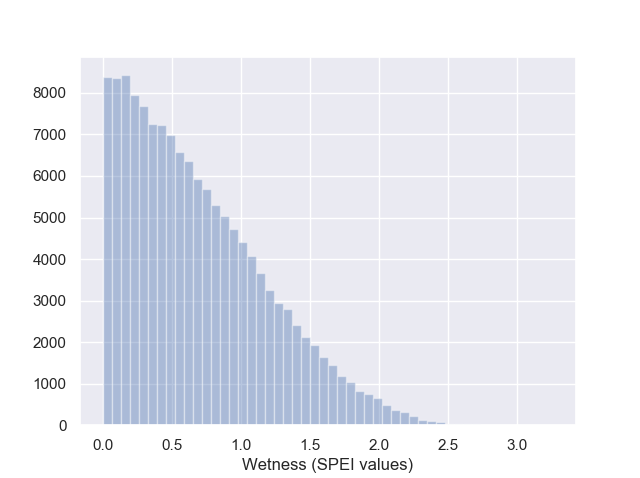}
\caption{Frequency Distribution of SPEI values at World Level for Dryness (left) and Wetness (right). Zero values are excluded.}
\label{fig:fig2}	
\end{figure}

\begin{figure}[ht]	
\centering
	\includegraphics[width=0.90\textwidth,height=1.1\textwidth]{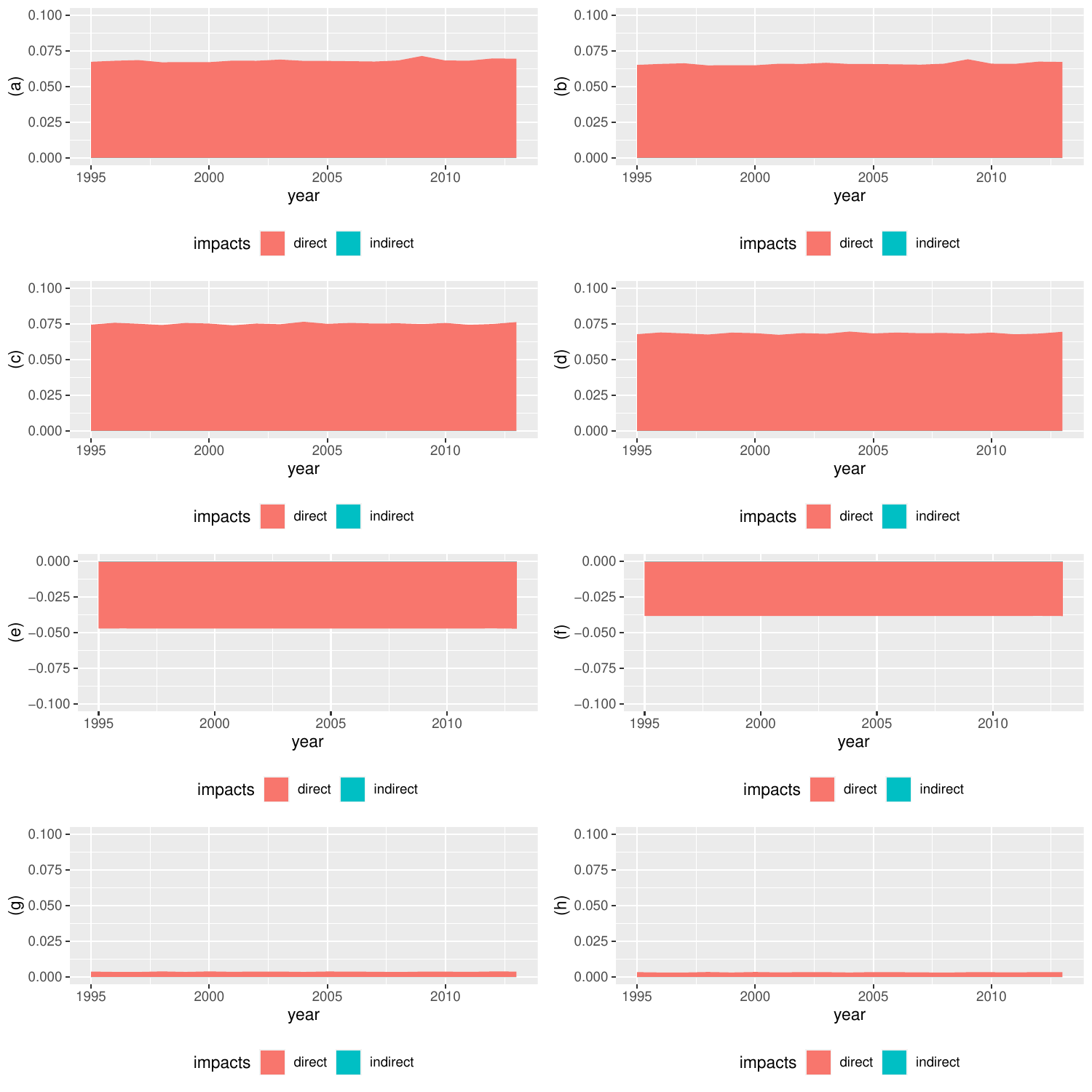}
        \caption{Time--varying Short--term (on the left) and Long--term (on the right) Total Marginal Effects from Model in eq. \eqref{eq:timediff} with respect to $GDP$ for (a-b) Europe, (c-d) South America, (e-f) South--East Asia, and (g-h) Africa, respectively. The Total Effects are split into the Direct (in pink) and Indirect (in green) Effects.}
\label{fig:tv_me}	
\end{figure}

\begin{figure}[ht]	
\centering
	\includegraphics[width=1\textwidth,height=0.9\textwidth]{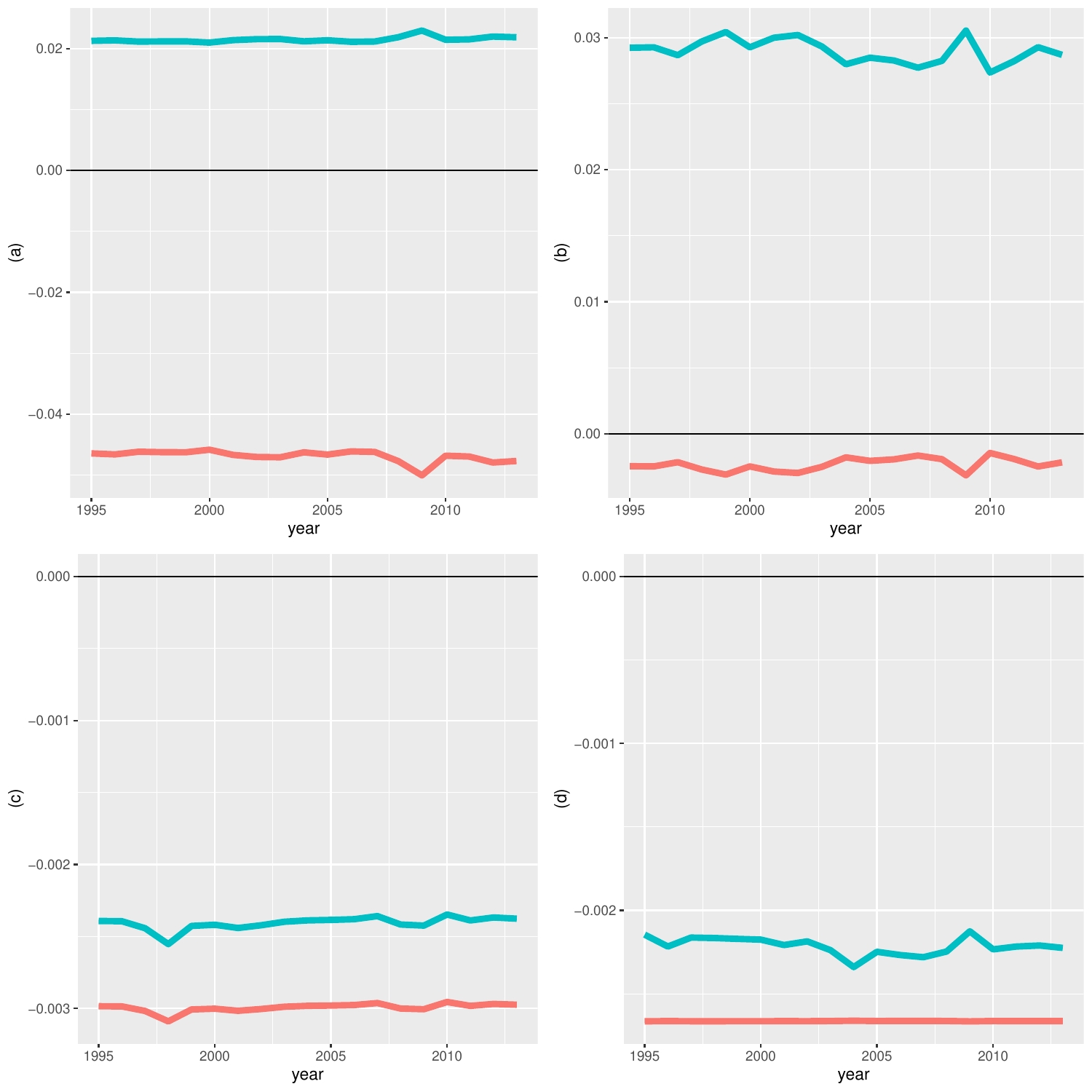}	
        \caption{Time--varying Marginal Effects from Model in eq. \eqref{eq:ecm} with respect to $Dryness$ (in red) and $Wetness$ (in green) for (a) Europe, (b) South America, (c) South--East Asia, and (d) Africa, respectively.}
\label{fig:tv_ecm_me}	
\end{figure}

\begin{figure}[ht]	
\centering
	\includegraphics[width=1\textwidth,height=0.50\textwidth]{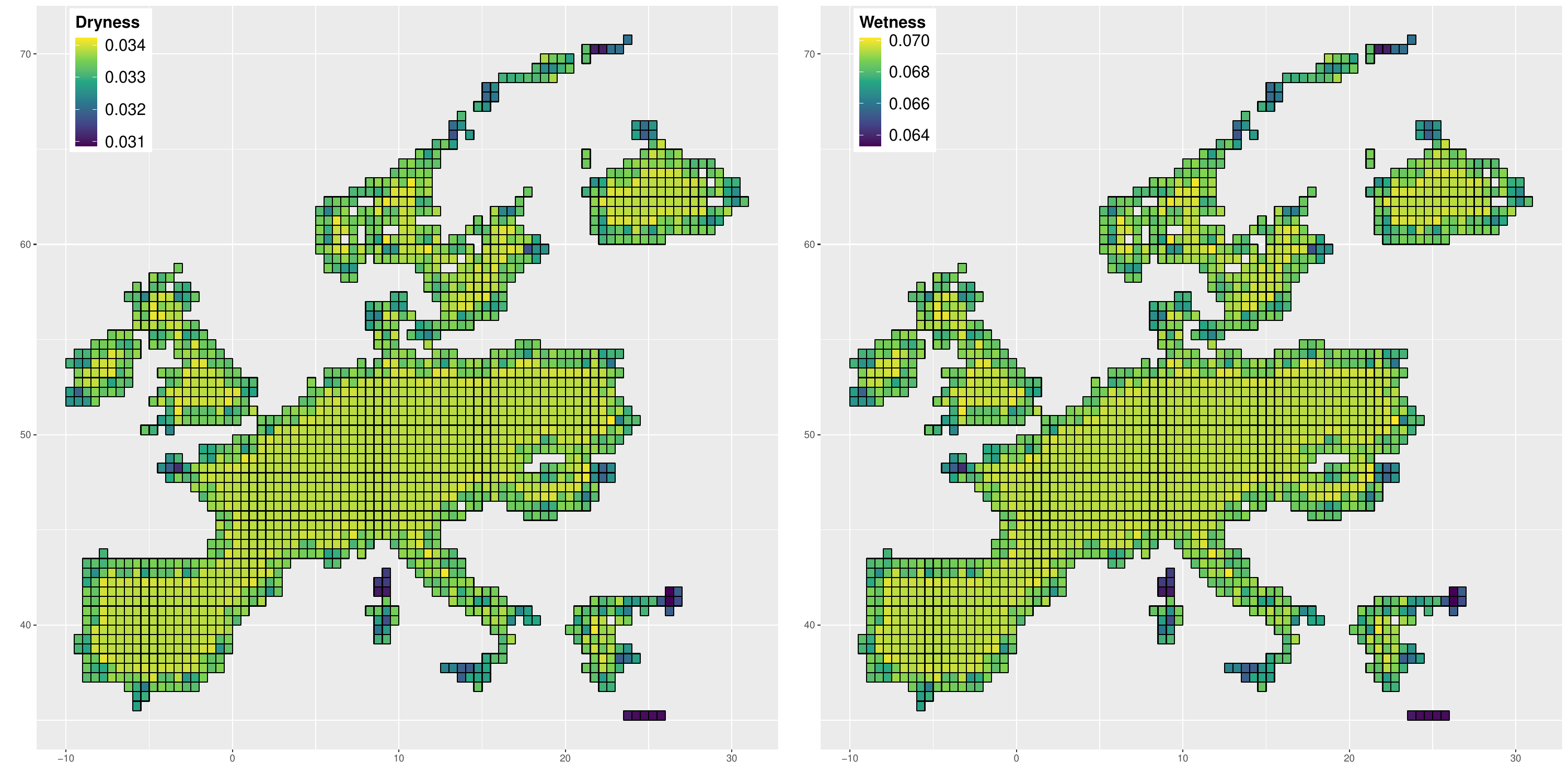}	         
        \caption{Local Short--term Indirect effects from model in eq. \eqref{eq:timediff} with respect to the variables $Dryness$ (on the left) and $Wetness$ (on the right) for Europe.}
\label{fig:fig4}	
\end{figure}

\begin{figure}[ht]	
\centering
	\includegraphics[width=1\textwidth,height=0.50\textwidth]{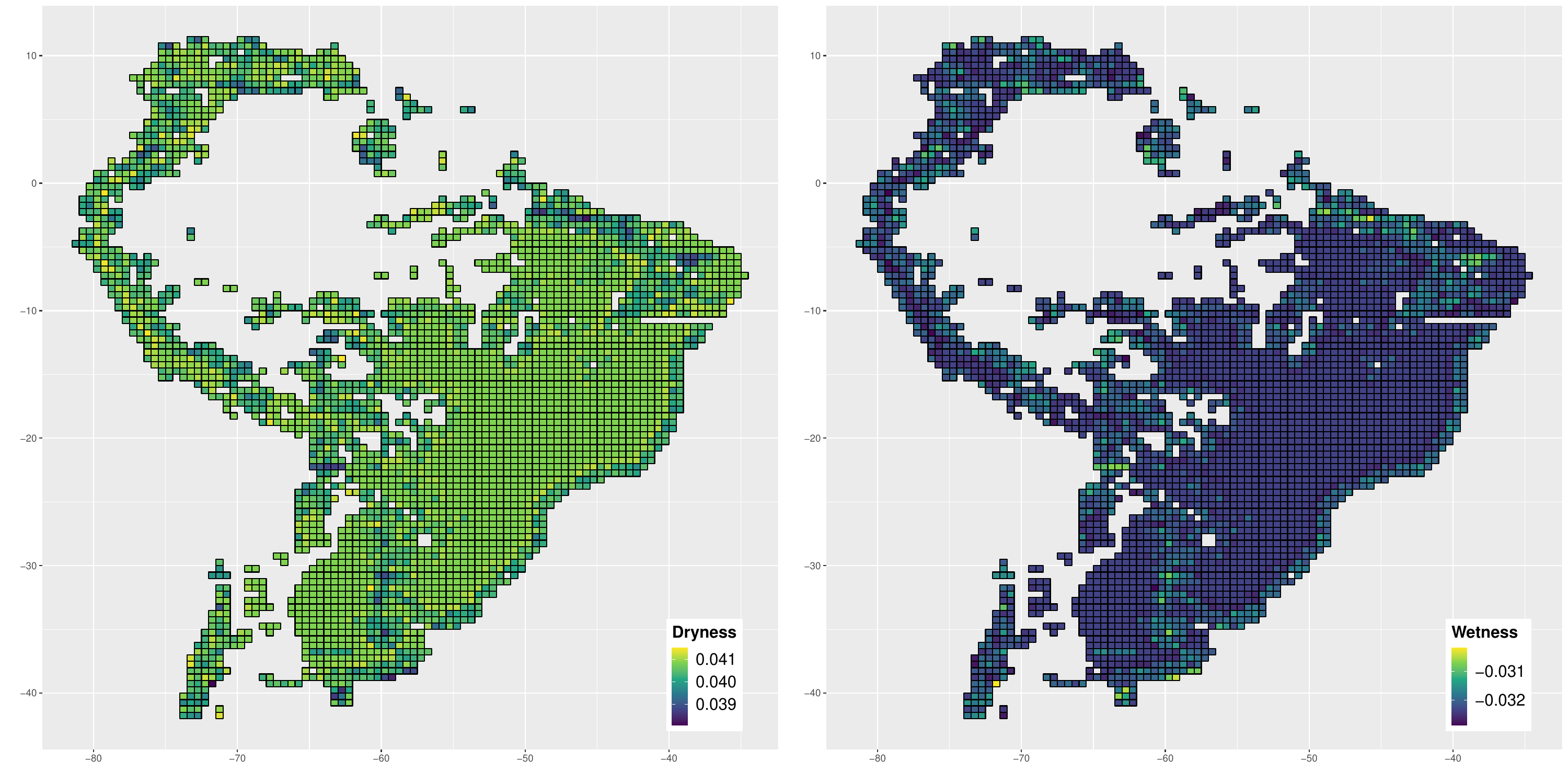}	
        \caption{Local Short--term Indirect effects from model in eq. \eqref{eq:timediff} with respect to the variables $Dryness$ (on the left) and $Wetness$ (on the right) for South America.}
\label{fig:fig5}	
\end{figure}

\begin{figure}[ht]	
\centering
	\includegraphics[width=1\textwidth,height=0.50\textwidth]{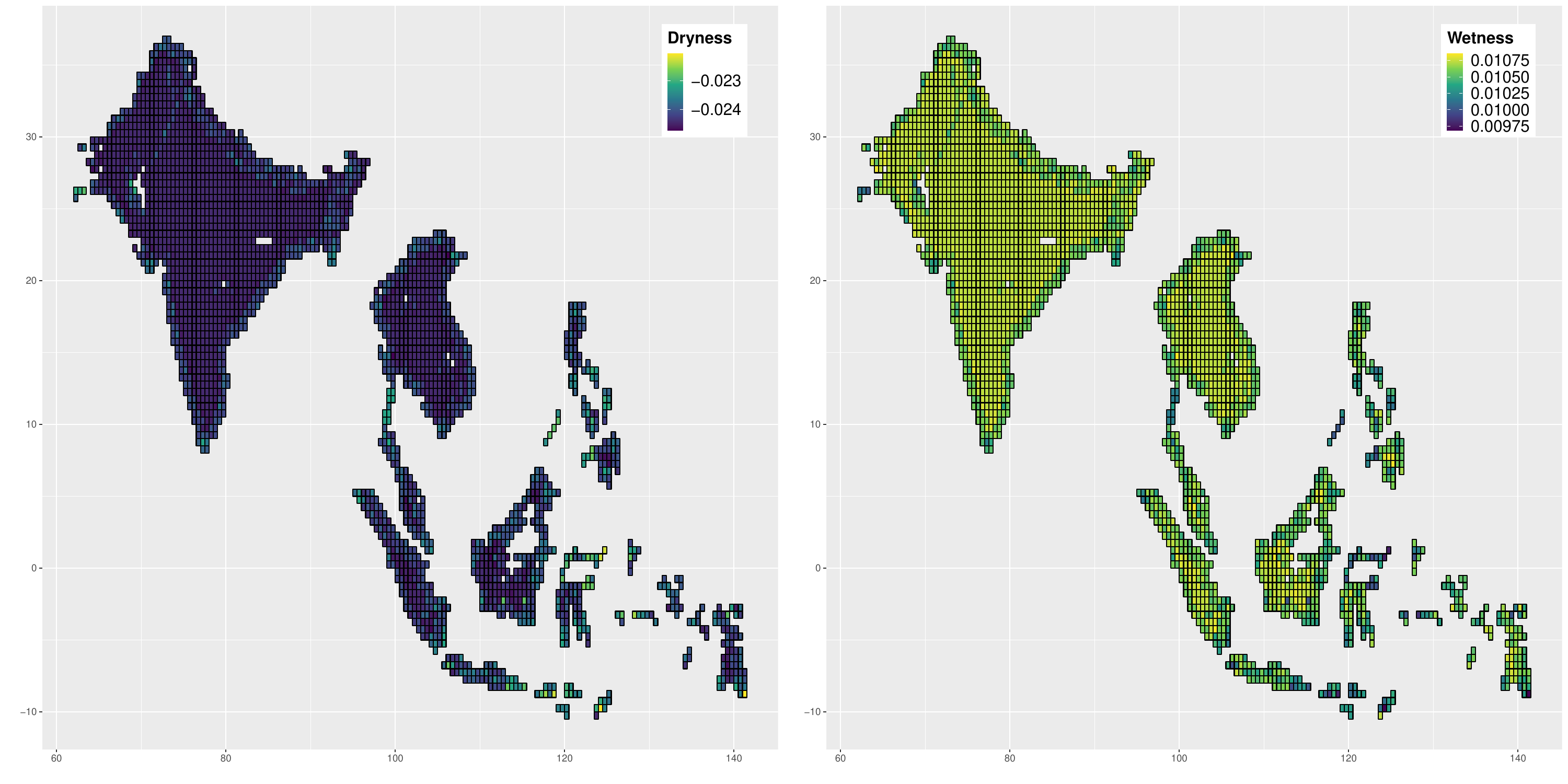}	
        \caption{Local Short--term Indirect effects from model in eq. \eqref{eq:timediff} with respect to the variables $Dryness$ (on the left) and $Wetness$ (on the right) for South--East Asia.}
\label{fig:fig6}	
\end{figure}

\begin{figure}[ht]	
\centering
	\includegraphics[width=1\textwidth,height=0.50\textwidth]{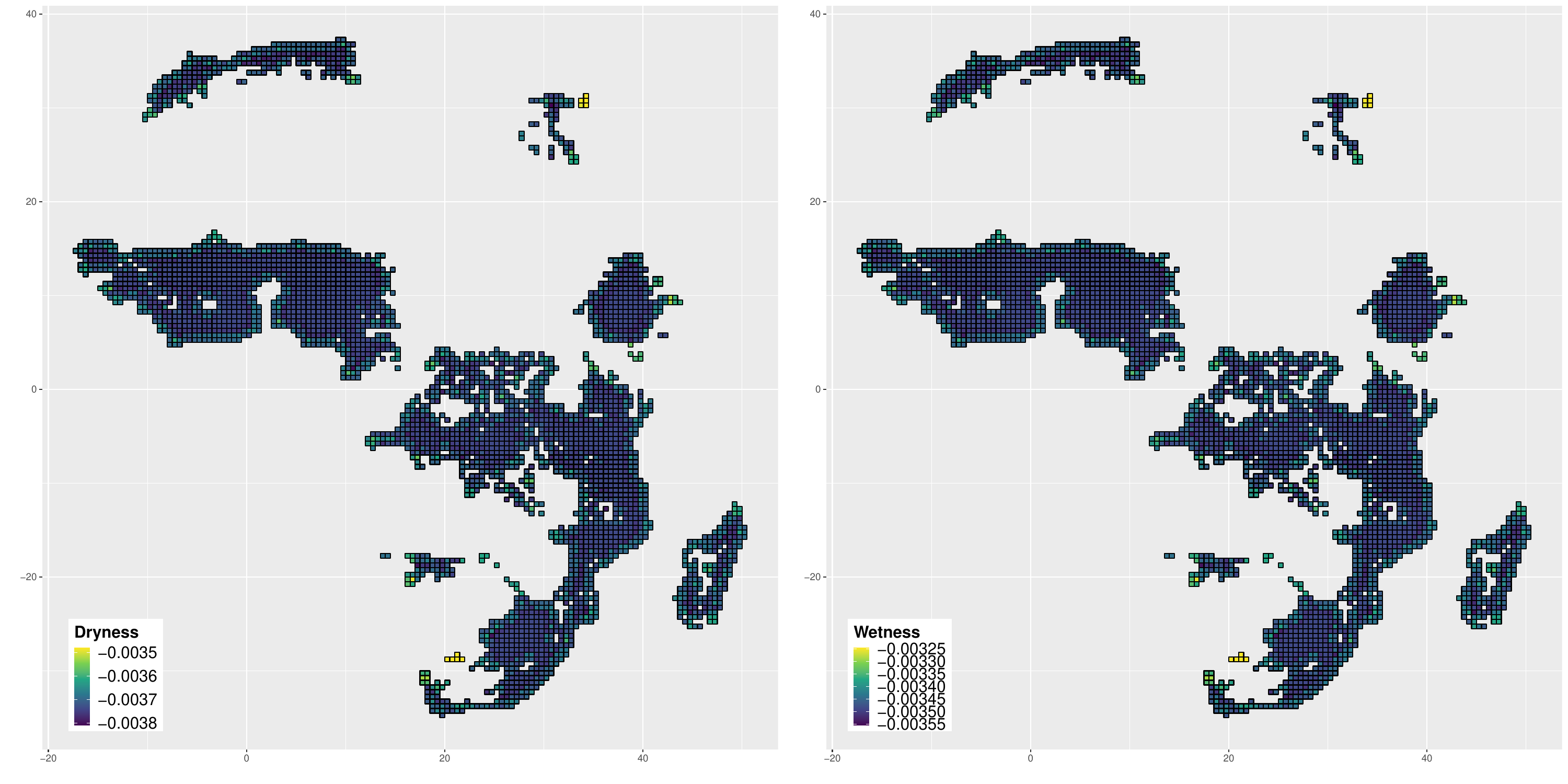}	
        \caption{Local Short--term Indirect effects from model in eq. \eqref{eq:timediff} with respect to the variables $Dryness$ (on the left) and $Wetness$ (on the right) for Africa.}
\label{fig:fig7}	
\end{figure}

Before considering estimation results, we detected the evolution of the Moran's I on the residuals of the model in equation \eqref{eq:timediff}, see Figure \ref{fig:moran}. As we can observe, Moran's I values fluctuate around the zero value of no autocorrelation in the error terms in Europe and South East Asia, while in South America and Africa there is a persistence of positive values, i.e. positive spatial error autocorrelations, over time. All the values are relatively low, ranging between $-0.10$ and $0.10$. The significant presence of error dependences in South America and Africa is also confirmed by the greater number of rejections, almost consecutive, of the null hypothesis of no autocorrelation, where I is not statistically different from 0. Although in some cases, especially in the period 2004--2013, we do not find the tests to be significant, we opted to consider spatial error autocorrelations in South America and Africa, since the majority of the null hypotheses are here rejected. 

Table \ref{tab:tab3c} shows the estimation results. First of all, it is interesting to note that, in South America and Africa, the estimates of the spatial error autocorrelations $\hat\lambda = \{0.848, 0.951\}$ are both very high and significant at $0.1\%$, reducing a bit the magnitude of the spatial autocorrelations in the dependent variables $\hat\rho = \{0.143, 0.257\}$ and confirming that potential unobserved shocks or factors could have a greater impact over time on fertilizer utilizations rather than social interactions among farmers. The price of Urea is also not significant both at time $t$ and $t-1$, probably also due to the adjustment of the standard errors when including relevant error autocorrelations, and revealing that no bias from omitted variable due to this price is at work here. Detailed comments on the other results are as follows.

Starting from Europe, we can see that the price of Urea is not significant neither at time $t$ nor at time $t-1$. It is possible that in advanced economies, such as the ones characterizing western Europe, farmers are scarcely sensitive to price variations, or else, the demand for a fundamental input such as nitrogen is rather price inelastic. An alternative explanation is that Urea is not the main source of nitrogen in this region being other sources such as Ammonia or Ammonium Nitrate preferred, and, despite the likely price correlation of substitute products, this last is not enough to reach statistical significance. It is possible to note that almost all contemporaneous variables become not significant including per capita GDP. The lagged term of wetness, instead, becomes significant compared to the regression without the price of Urea, testifying that also in Europe, as was for South America, there may be an increase of fertilization to counteract the possible run--off of nitrogen after a wet season. 

Also in South America the effect of fertilizer price is not significant neither at time $t$ nor at time $t-1$. The same reasons seen for Europe may well apply here. It must be further noted that the per capita GDP coefficients have the same signs as the ones of Europe in Table \ref{tab:tab3}, suggesting that South America could be in the concave section of the Kuznets curve rather than in its ascending part. Another interesting fact to notice is the change of sign for the lagged coefficient of the agricultural output price, now become positive (and significant at 10\% level). The negative sign for this variable has remained only in Africa, but with reduced significance whereas in South East Asia has lost significance. Finally, it must be noted the loss of significance of the lagged weather conditions.

In South East Asia the fertilizer price effect is statistically strong and negative both at time $t$ and $t-1$. This implies that farmers are very price sensitive in their choice to fertilize the soil. Compared to the regression with the price of Urea omitted, it is possible to observe the loss of significance of per capita GDP and a gain of significance for the effect of contemporaneous wetness, with it being positive. The same variable gains significance also in Africa together with the contemporaneous effect of dryness, that, however, has a negative sign. This may imply that farmers tend to fertilize more when they observe or foresee abundant rains whereas they avoid to do so in the opposite case, possibly for saving inputs. This practice, however, can be very deleterious given the higher need of nitrogen under dry conditions mentioned earlier.

These are the most significant variations produced by the introduction of the price of Urea as a proxy for the cost of nitrogen. The coefficients of the other variables are, overall, similar to the ones in Table \ref{tab:tab3}. The availability of better data for the cost of nitrogen could improve the analysis thus leaving room for further studies once there will be such availability.  
\subsection{Yeo--Johnson Power Transformation and Robustness Checks}
\label{sec:robust}
In this Section we briefly report the main results of some robustness checks of our model specification in equation \eqref{eq:timediff}, i.e. the estimation results are almost the same in terms of both the sign and the magnitude for the majority of the regressors considered. Tables on both the regression results after YJ transformation and the other robustness checks are reported in the Supplementary material of this paper.

First of all, the estimated parameter values are robust to the Yeo--Johnson (YJ) transformation, which differently from the Box--Cox transformation allows also for zero values of $\by$. The values of the shift parameter $\lambda$ are equal to $\{0.55; 0.3; 0.4; -1.2\}$ for Europa, South America, South East Asia and Africa, respectively, see Figure \ref{fig:fig1}, with $\lambda=1$ means no transformation. Only slight differences can be found, especially for the spatial autoregressive parameter $\rho$ in those macro--regions in which a more heavily data transformation has been applied.
Secondly, as expected, the results are also robust to different sparse weighting matrices. Specifically, we consider $k=\{11, 18\}$ for the $k$--nn approach and the queen scheme, for which the weighting matrix is still largely sparse. Only slight differences can be found in the magnitude of the spatial autoregressive parameter $\rho$.

We also found that the inclusion of the spatially lagged regressors $\sW\bX$ for GDP, Dryness and Wetness, can be in some situations, and as expected, statistically significant. In particular, only the estimated coefficients of GDP, Dryness and Wetness at time $t$ are affected by the inclusion of the same spatially lagged regressors, since, obviously, they can be highly correlated among each other. However, their inclusion do not change substantially the sign and the magnitude of the other estimated parameters, avoiding in this specific situation a heavy omitted variable bias when they are excluded. 
Finally, the model in equation \eqref{eq:timediff} can be sometimes over--specified when including simultaneously the terms $\sW\by_t$, $\sW\by_{t-1}$ and $\by_{t-1}$. We then excluded the term $\sW\by_{t-1}$, finding again no substantial differences.
\section{Conclusions}
\label{sec:conclusion}
The present paper analyse the relation between abnormal weather conditions and fertilizer applications at world level, by considering four macro--regions -- Europe (CAP), South America, South--East Asia and Africa -- and using a recent dataset of gridded data which covers more than 20 years (1993--2013). The selected exogenous covariates include dryness and wetness (excess of rain) as indicators of weather conditions, derived from the SPEI index, as well as per--capita GDP (PPP and constant 2011 dollars). Furthermore, the price of agricultural output (PAO) and a proxy for the price of fertilizer (PF, i.e. price of Urea) are included to avoid potential omitted variable bias problems. Different menthods of data interpolation are adopted to fill some missing values.

The methodological strategy is based on the use of a spatial dynamic panel data (SDPD) model that deals with both space and time dependence effects, i.e. spatial (cross--sectional) and serial correlations, as well as both space (individual) and time fixed effects. To avoid inconsistency due to potential non--stability of the SDPD model, we conduct Wald tests on the null hypothesis of spatial cointegration and we then time first--differenciate the model specification. Moreover, we also calculate time--varying Moran's I indeces and tests to control for potential (residual) spatial error autocorrelations. Finally, some robustness checks like regression results with different weighting schemes, with the YJ power trasformation of the dependent variables, and with or without some additional covariates have been also included.

The main results are as follows. Both the dynamic and the spatial dependence parameters are statistically very strong with the latter being positive and the former negative for all the macro--regions. This testifies both certain behavioural similarities in agronomic practices among neighbouring areas, probably due to similar agro--climatic conditions, and the presence of knowledge deriving from the past. The spatio--temporal coefficient is also strongly significant, and positive, in all the macro--regions, even when considering the error correlation. This validates the choice of the spatial dynamic model, revealing that not only pure spatial and temporal dependence parameters play an important role, but also the spatio--temporal coefficient could hide spillover effects, which will be shown one year later. Looking at the climatic variables, it is possible to observe a rather differentiated response to weather anomalies. In particular, whereas in Europe and in South America present and past levels of dryness and wetness are generally significant predictors of N--fertilization, the opposite is true in South--East Asia and Africa. This may imply a lower capacity, possibly due to knowledge or technical deficiencies, to adapt fertilization to abnormal weather conditions. Not surprisingly, the area where good agronomic practices seem to be observed more often is Europe, whereas South America shows some contrasting results. The analysis of per--capita GDP suggests the existence of an environmental Kuznets curve for N--fertilizer in Europe, and also in South America when including PF, showing an inverted U--shape relation between per--capita GDP and the amount of applied fertilizer. South--East Asia and Africa, instead, display a linearly positive relation between fertilization and the time lag of per--capita GDP, suggesting the possibility that they are still in the ascending section of the Kuznets curve. This last element may also be interpreted as a sign of the presence of poverty traps, thus calling for ad hoc policy interventions.

Given the recent high relevance of forecasting with spatio--temporal models and the raising availability of spatio--temporal data, it is finally worth mentioning that this paper could lay the foundations for correctly specifying the model specification used as basis for time predictions of fertilization worldwide. This project should deserve the right attention for future research. 
\section*{Acknowledgements}
%
%
\clearpage
%
\bibliographystyle{apalike}
\bibliography{bib_fert}

\begin{thebibliography}{}

\bibitem[Anselin et~al., 2008]{anselin2008spatial}
Anselin, L., Le~Gallo, J., and Jayet, H. (2008).
\newblock Spatial panel econometrics.
\newblock In {\em The econometrics of panel data}, pages 625--660. Springer.

\bibitem[Baltagi et~al., 2003]{baltagi2003testing}
Baltagi, B.~H., Song, S.~H., and Koh, W. (2003).
\newblock Testing panel data regression models with spatial error correlation.
\newblock {\em Journal of econometrics}, 117(1):123--150.

\bibitem[Barrett, 2007]{barrett07}
Barrett, C.~B. (2007).
\newblock Displaced distortions: Financial market failures and seemingly
  inefficient resource allocation in low-income rural communities.
\newblock {\em Development economics between markets and institutions:
  Incentives for growth, food security and sustainable use of the environment},
  pages 73--86.

\bibitem[Baylis et~al., 2011]{baylis2011spatial}
Baylis, K., Paulson, N.~D., and Piras, G. (2011).
\newblock Spatial approaches to panel data in agricultural economics: a climate
  change application.
\newblock {\em Journal of Agricultural and Applied Economics}, 43(3):325--338.

\bibitem[Belotti et~al., 2017]{belotti2017spatial}
Belotti, F., Hughes, G., and Mortari, A.~P. (2017).
\newblock Spatial panel-data models using stata.
\newblock {\em The Stata Journal}, 17(1):139--180.

\bibitem[Campbell, 1999]{campbell99}
Campbell, D.~J. (1999).
\newblock
  \href{https://link.springer.com/article/10.1023/A:1018789623581}{Response to
  drought among farmers and herders in southern Kajiado District, Kenya: A
  comparison of 1972-1976 and 1994-1995}.
\newblock {\em Human ecology}, 27(3):377--416.

\bibitem[Debarsy et~al., 2012]{debarsy2012interpreting}
Debarsy, N., Ertur, C., and LeSage, J.~P. (2012).
\newblock Interpreting dynamic space--time panel data models.
\newblock {\em Statistical Methodology}, 9(1-2):158--171.

\bibitem[Devereux, 2007]{devereux07}
Devereux, S. (2007).
\newblock
  \href{https://onlinelibrary.wiley.com/doi/abs/10.1111/j.1574-0862.2007.00234.x}{The
  impact of droughts and floods on food security and policy options to
  alleviate negative effects}.
\newblock {\em Agricultural Economics}, 37:47--58.

\bibitem[Ding et~al., 2009]{ding09}
Ding, Y., Schoengold, K., and Tadesse, T. (2009).
\newblock
  \href{https://www.jstor.org/stable/41548424?seq=1#metadata_info_tab_contents}{The
  impact of weather extremes on agricultural production methods: Does drought
  increase adoption of conservation tillage practices?}
\newblock {\em Journal of Agricultural and Resource Economics}, pages 395--411.

\bibitem[Elhorst, 2014]{elhorst2014spatial}
Elhorst, J.~P. (2014).
\newblock {\em Spatial Econometrics: From Cross-sectional Data to Spatial
  Panels}, volume 479.
\newblock Springer.

\bibitem[Elhorst et~al., 2013]{elhorst2013impact}
Elhorst, P., Zandberg, E., and De~Haan, J. (2013).
\newblock The impact of interaction effects among neighbouring countries on
  financial liberalization and reform: A dynamic spatial panel data approach.
\newblock {\em Spatial Economic Analysis}, 8(3):293--313.

\bibitem[Fischer and Knutti, 2015]{fischer15}
Fischer, E.~M. and Knutti, R. (2015).
\newblock
  \href{https://www.nature.com/articles/nclimate2617#ref4}{Anthropogenic
  contribution to global occurrence of heavy-precipitation and high-temperature
  extremes}.
\newblock {\em Nature Climate Change}, 5(6):560.

\bibitem[Fisher et~al., 2015]{fisher15}
Fisher, M., Abate, T., Lunduka, R.~W., Asnake, W., Alemayehu, Y., and Madulu,
  R.~B. (2015).
\newblock
  \href{https://link.springer.com/article/10.1007/s10584-015-1459-2}{Drought
  tolerant maize for farmer adaptation to drought in sub-Saharan Africa:
  Determinants of adoption in eastern and southern Africa}.
\newblock {\em Climatic Change}, 133(2):283--299.

\bibitem[Hartmann et~al., 2011]{hartmann11}
Hartmann, A.~A., Buchmann, N., and Niklaus, P.~A. (2011).
\newblock \href{https://link.springer.com/article/10.1007/s11104-010-0690-x}{A
  study of soil methane sink regulation in two grasslands exposed to drought
  and N fertilization}.
\newblock {\em Plant and soil}, 342(1-2):265--275.

\bibitem[Hossain et~al., 2016]{hossain16}
Hossain, M.~N., Chowdhury, S., and Paul, S.~K. (2016).
\newblock
  \href{https://link.springer.com/article/10.1007/s11069-016-2360-7}{Farmer-level
  adaptation to climate change and agricultural drought: empirical evidences
  from the Barind region of Bangladesh}.
\newblock {\em Natural Hazards}, 83(2):1007--1026.

\bibitem[Iizumi and Ramankutty, 2015]{iizumi15}
Iizumi, T. and Ramankutty, N. (2015).
\newblock
  \href{https://www.sciencedirect.com/science/article/pii/S2211912414000583}{How
  do weather and climate influence cropping area and intensity?}
\newblock {\em Global Food Security}, 4:46--50.

\bibitem[Lee and Yu, 2010a]{lee2010some}
Lee, L.-f. and Yu, J. (2010a).
\newblock Some recent developments in spatial panel data models.
\newblock {\em Regional Science and Urban Economics}, 40(5):255--271.

\bibitem[Lee and Yu, 2010b]{lee2010spatial}
Lee, L.-F. and Yu, J. (2010b).
\newblock A spatial dynamic panel data model with both time and individual
  fixed effects.
\newblock {\em Econometric Theory}, 26(2):564--597.

\bibitem[Lee and Yu, 2011]{lee2011estimation}
Lee, L.-f. and Yu, J. (2011).
\newblock {\em Estimation of spatial panels}.
\newblock Now Publishers Inc.

\bibitem[Lee and Yu, 2014]{lee2014efficient}
Lee, L.-f. and Yu, J. (2014).
\newblock Efficient gmm estimation of spatial dynamic panel data models with
  fixed effects.
\newblock {\em Journal of Econometrics}, 180(2):174--197.

\bibitem[Lee and Yu, 2020]{lee2020initial}
Lee, L.-f. and Yu, J. (2020).
\newblock Initial conditions of dynamic panel data models: on within and
  between equations.
\newblock {\em The Econometrics Journal}, 23(1):115--136.

\bibitem[Lesk et~al., 2016]{lesk16}
Lesk, C., Rowhani, P., and Ramankutty, N. (2016).
\newblock \href{https://www.nature.com/articles/nature16467}{Influence of
  extreme weather disasters on global crop production}.
\newblock {\em Nature}, 529(7584):84.

\bibitem[Li et~al., 2016]{li16}
Li, F., Dong, S., Li, F., and Yang, L. (2016).
\newblock \href{https://link.springer.com/article/10.1007/s11783-014-0700-y}{Is
  there an inverted U-shaped curve? Empirical analysis of the environmental
  Kuznets curve in agrochemicals}.
\newblock {\em Frontiers of Environmental Science \& Engineering},
  10(2):276--287.

\bibitem[Li and Yang, 2020]{li2020spatial}
Li, L. and Yang, Z. (2020).
\newblock Spatial dynamic panel data models with correlated random effects.
\newblock {\em Journal of Econometrics}.

\bibitem[Lobell and Field, 2007]{lobell07}
Lobell, D.~B. and Field, C.~B. (2007).
\newblock
  \href{http://iopscience.iop.org/article/10.1088/1748-9326/2/1/014002/meta}{Global
  scale climate--crop yield relationships and the impacts of recent warming}.
\newblock {\em Environmental research letters}, 2(1):014002.

\bibitem[Lu and Tian, 2017]{lu17}
Lu, C. and Tian, H. (2017).
\newblock \href{https://www.earth-syst-sci-data.net/9/181/2017/}{Global
  nitrogen and phosphorus fertilizer use for agriculture production in the past
  half century: shifted hot spots and nutrient imbalance}.
\newblock {\em Earth System Science Data}, 9(1):181--192.

\bibitem[McKee et~al., 1993]{mckee93}
McKee, T.~B., Doesken, N.~J., Kleist, J., et~al. (1993).
\newblock
  \href{https://climate.colostate.edu/pdfs/relationshipofdroughtfrequency.pdf}{The
  relationship of drought frequency and duration to time scales}.
\newblock In {\em Proceedings of the 8th Conference on Applied Climatology},
  volume~17, pages 179--183. American Meteorological Society Boston, MA.

\bibitem[Millo, 2017]{millo2017simple}
Millo, G. (2017).
\newblock A simple randomization test for spatial correlation in the presence
  of common factors and serial correlation.
\newblock {\em Regional Science and Urban Economics}, 66:28--38.

\bibitem[Millo et~al., 2012]{millo2012splm}
Millo, G., Piras, G., et~al. (2012).
\newblock splm: Spatial panel data models in r.
\newblock {\em Journal of Statistical Software}, 47(1):1--38.

\bibitem[Moran, 1950]{moran1950notes}
Moran, P.~A. (1950).
\newblock Notes on continuous stochastic phenomena.
\newblock {\em Biometrika}, 37(1/2):17--23.

\bibitem[Morton et~al., 2015]{morton15}
Morton, L.~W., Hobbs, J., Arbuckle, J.~G., and Loy, A. (2015).
\newblock
  \href{https://dl.sciencesocieties.org/publications/jeq/abstracts/44/3/810}{Upper
  Midwest climate variations: Farmer responses to excess water risks}.
\newblock {\em Journal of environmental quality}, 44(3):810--822.

\bibitem[Mutl and Pfaffermayr, 2011]{mutl2011hausman}
Mutl, J. and Pfaffermayr, M. (2011).
\newblock The hausman test in a cliff and ord panel model.
\newblock {\em The Econometrics Journal}, 14(1):48--76.

\bibitem[Nash et~al., 2011]{nash2011unifying}
Nash, J.~C., Varadhan, R., et~al. (2011).
\newblock Unifying optimization algorithms to aid software system users: optimx
  for r.
\newblock {\em Journal of Statistical Software}, 43(9):1--14.

\bibitem[Pall et~al., 2011]{pall11}
Pall, P., Aina, T., Stone, D.~A., Stott, P.~A., Nozawa, T., Hilberts, A.~G.,
  Lohmann, D., and Allen, M.~R. (2011).
\newblock \href{https://www.nature.com/articles/nature09762}{Anthropogenic
  greenhouse gas contribution to flood risk in England and Wales in autumn
  2000}.
\newblock {\em Nature}, 470(7334):382.

\bibitem[Palmer, 1965]{palmer65}
Palmer, W.~C. (1965).
\newblock Meteorological drought. research paper no. 45. washington, dc: Us
  department of commerce.
\newblock {\em Weather Bureau}, page~59.

\bibitem[Pandey et~al., 2007]{pandey07}
Pandey, S., Bhandari, H.~S., and Hardy, B. (2007).
\newblock {\em Economic costs of drought and rice farmers' coping mechanisms: a
  cross-country comparative analysis}.
\newblock Int. Rice Res. Inst.

\bibitem[Parent and LeSage, 2012]{parent2012spatial}
Parent, O. and LeSage, J.~P. (2012).
\newblock Spatial dynamic panel data models with random effects.
\newblock {\em Regional Science and Urban Economics}, 42(4):727--738.

\bibitem[Potter et~al., 2010]{potter10}
Potter, P., Ramankutty, N., Bennett, E.~M., and Donner, S.~D. (2010).
\newblock
  \href{https://journals.ametsoc.org/doi/full/10.1175/2009EI288.1}{Characterizing
  the spatial patterns of global fertilizer application and manure production}.
\newblock {\em Earth Interactions}, 14(2):1--22.

\bibitem[Purcell et~al., 2004]{purcell04}
Purcell, L.~C., Serraj, R., Sinclair, T.~R., and De, A. (2004).
\newblock
  \href{https://dl.sciencesocieties.org/publications/cs/abstracts/44/2/484}{Soybean
  N 2 fixation estimates, ureide concentration, and yield responses to
  drought}.
\newblock {\em Crop Science}, 44(2):484--492.

\bibitem[Rosenzweig et~al., 2002]{rosenzweig02}
Rosenzweig, C., Tubiello, F.~N., Goldberg, R., Mills, E., and Bloomfield, J.
  (2002).
\newblock
  \href{https://www.sciencedirect.com/science/article/abs/pii/S0959378002000080}{Increased
  crop damage in the US from excess precipitation under climate change}.
\newblock {\em Global Environmental Change}, 12(3):197--202.

\bibitem[Shi and Lee, 2017]{shi2017spatial}
Shi, W. and Lee, L.-f. (2017).
\newblock Spatial dynamic panel data models with interactive fixed effects.
\newblock {\em Journal of Econometrics}, 197(2):323--347.

\bibitem[Soule et~al., 2000]{soule00}
Soule, M.~J., Tegene, A., and Wiebe, K.~D. (2000).
\newblock
  \href{https://onlinelibrary.wiley.com/doi/abs/10.1111/0002-9092.00097}{Land
  tenure and the adoption of conservation practices}.
\newblock {\em American journal of agricultural economics}, 82(4):993--1005.

\bibitem[Stott et~al., 2004]{stott04}
Stott, P.~A., Stone, D.~A., and Allen, M.~R. (2004).
\newblock \href{https://www.nature.com/articles/nature03089}{Human contribution
  to the European heatwave of 2003}.
\newblock {\em Nature}, 432(7017):610.

\bibitem[{The World Bank}, 2018a]{WBruremp17}
{The World Bank} (2018a).
\newblock \href{https://data.worldbank.org/indicator/SL.AGR.EMPL.ZS}{Employment
  in agriculture (\% of total employment)}.
\newblock Accessed on 20/09/2018 at
  https://data.worldbank.org/indicator/SL.AGR.EMPL.ZS.

\bibitem[{The World Bank}, 2018b]{WBrurpop17}
{The World Bank} (2018b).
\newblock \href{https://data.worldbank.org/indicator/SP.RUR.TOTL.ZS}{Rural
  population (\% of total population)}.
\newblock Accessed on 20/09/2018 at
  https://data.worldbank.org/indicator/SP.RUR.TOTL.ZS.

\bibitem[T{\'o}th et~al., 2014]{toth14}
T{\'o}th, G., Guicharnaud, R.-A., T{\'o}th, B., and Hermann, T. (2014).
\newblock
  \href{https://www.sciencedirect.com/science/article/pii/S1161030113001950}{Phosphorus
  levels in croplands of the European Union with implications for P fertilizer
  use}.
\newblock {\em European Journal of Agronomy}, 55:42--52.

\bibitem[Van~Metre et~al., 2016]{van16}
Van~Metre, P.~C., Frey, J.~W., Musgrove, M., Nakagaki, N., Qi, S., Mahler,
  B.~J., Wieczorek, M.~E., and Button, D.~T. (2016).
\newblock
  \href{https://dl.sciencesocieties.org/publications/jeq/abstracts/45/5/1696}{High
  nitrate concentrations in some Midwest United States streams in 2013 after
  the 2012 drought}.
\newblock {\em Journal of Environmental Quality}, 45(5):1696--1704.

\bibitem[Vicente-Serrano et~al., 2010]{vicente10}
Vicente-Serrano, S.~M., Beguer{\'\i}a, S., and L{\'o}pez-Moreno, J.~I. (2010).
\newblock \href{https://journals.ametsoc.org/doi/abs/10.1175/2009JCLI2909.1}{A
  multiscalar drought index sensitive to global warming: the standardized
  precipitation evapotranspiration index}.
\newblock {\em Journal of climate}, 23(7):1696--1718.

\bibitem[Vlek and Byrnes, 1986]{vlek86}
Vlek, P.~L. and Byrnes, B.~H. (1986).
\newblock
  \href{https://link.springer.com/chapter/10.1007/978-94-009-4428-2_7}{The
  efficacy and loss of fertilizer N in lowland rice}.
\newblock In {\em Nitrogen economy of flooded rice soils}, pages 131--147.
  Springer.

\bibitem[Wells et~al., 2004]{wells04}
Wells, N., Goddard, S., and Hayes, M.~J. (2004).
\newblock
  \href{https://journals.ametsoc.org/doi/abs/10.1175/1520-0442(2004)017{\%}3C2335:ASPDSI{\%}3E2.0.CO;2}{A
  self-calibrating Palmer drought severity index}.
\newblock {\em Journal of Climate}, 17(12):2335--2351.

\bibitem[Wooldridge, 2010]{wooldridge2010econometric}
Wooldridge, J.~M. (2010).
\newblock {\em Econometric analysis of cross section and panel data}.
\newblock MIT press.

\bibitem[Xu and Lee, 2015]{xu2015maximum}
Xu, X. and Lee, L.-f. (2015).
\newblock Maximum likelihood estimation of a spatial autoregressive tobit
  model.
\newblock {\em Journal of Econometrics}, 188(1):264--280.

\bibitem[Yeo and Johnson, 2000]{yeo2000new}
Yeo, I.-K. and Johnson, R.~A. (2000).
\newblock A new family of power transformations to improve normality or
  symmetry.
\newblock {\em Biometrika}, 87(4):954--959.

\bibitem[Yu et~al., 2008]{yu2008quasi}
Yu, J., De~Jong, R., and Lee, L.-f. (2008).
\newblock Quasi-maximum likelihood estimators for spatial dynamic panel data
  with fixed effects when both n and t are large.
\newblock {\em Journal of Econometrics}, 146(1):118--134.

\bibitem[Yu et~al., 2012]{yu2012estimation}
Yu, J., de~Jong, R., and Lee, L.-f. (2012).
\newblock Estimation for spatial dynamic panel data with fixed effects: The
  case of spatial cointegration.
\newblock {\em Journal of Econometrics}, 167(1):16--37.

\bibitem[Zipper et~al., 2016]{zipper16}
Zipper, S.~C., Qiu, J., and Kucharik, C.~J. (2016).
\newblock
  \href{https://iopscience.iop.org/article/10.1088/1748-9326/11/9/094021/meta}{Drought
  effects on US maize and soybean production: spatiotemporal patterns and
  historical changes}.
\newblock {\em Environmental Research Letters}, 11(9):094021.

\end{thebibliography}
\clearpage
\appendix
\section{Local Marginal Effects}
\label{sec:local_me}
Here we briefly show how to consider local marginal effects, i.e. unit--specific marginal effects. In this paper we consider the spatial heterogeneity in terms of the time--invariant short--term marginal effects in eq. \eqref{eq:marginal_short}, with respect to the variables $DRY_t$ ($\x_{2,t}$) and $WET_t$ ($\x_{3,t}$). Alternatively, if they are time--varying, one can consider to show specific marginal effects in one point in time, say $t$. Let $\x_{k,t}$ be a covariate of interest at time $t$ and $\by_{t}$ the dependent variable at time $t$. We are then interested in defining the marginal effect of $\x_{k,t}$ on $\by_{t}$ for each spatial unit $i = 1, \dots, n$. Therefore we have
\begin{equation*}
\Large
\begin{bmatrix}
\frac{\partial y_{1,t}}{\partial x_{1k,t}} & \frac{\partial y_{1,t}}{\partial x_{2k,t}} & \dots & \frac{\partial y_{1,t}}{\partial x_{ik,t}} & \dots & \frac{\partial y_{1,t}}{\partial x_{nk,t}} \\
\frac{\partial y_{2,t}}{\partial x_{1k,t}} & \frac{\partial y_{2,t}}{\partial x_{2k,t}} & \dots & \frac{\partial y_{2,t}}{\partial x_{ik,t}} & \dots & \frac{\partial y_{2,t}}{\partial x_{nk,t}} \\
\vdots & \vdots & \ddots & \dots & \dots & \dots \\
\vdots & \vdots & \vdots & \frac{\partial y_{i,t}}{\partial x_{ik,t}} & \dots & \dots \\
\vdots & \vdots & \vdots & \vdots & \ddots & \dots \\
\frac{\partial y_{n,t}}{\partial x_{1k,t}} & \frac{\partial y_{n,t}}{\partial x_{2k,t}} & \dots & \frac{\partial y_{n,t}}{\partial x_{ik,t}} & \dots & \frac{\partial y_{n,t}}{\partial x_{nk,t}}
\end{bmatrix}
\end{equation*}
the elements of which are direct effects (diagonal elements) and the indirect effects (cross--diagonal elements). However, we should note that the information coming from the indirect effects, i.e. the average of the off--diagonal elements, is different according to considering the lower or the upper triangular matrix. Indeed, for instance, in the first row we recognize the indirect effects of all the variables $x_{1k,t}, \dots, x_{nk,t}$ on the dependent variable $y_{1,t}$, while in the first column we observe the indirect effects of the variable $x_{1k,t}$ on all the dependent variables $y_{1,t}, \dots, y_{n,t}$, excluding the first element (the direct effect) in both cases. In our paper we consider the first case of heterogeneity for each $y_{i,t}$, i.e. looking at each row, and then summarise the information by taking row--sums. The results are shown in Figures \ref{fig:fig4}, \ref{fig:fig5}, \ref{fig:fig6}, and \ref{fig:fig7}.
\clearpage
%
%
\section{Tables}
\begin{table}[htbp!]
	\centering
	\caption{Summary Statistics over the entire Panel dimension.}
	\resizebox{0.7\textwidth}{0.43\textheight}{
	\begin{tabular}{lcccccc}
		\hline\hline
		\multicolumn{1}{c}{\textbf{Variables}} & \textbf{Mean} & \textbf{Median} & \textbf{Std. Dev.} & \textbf{Min.} & \textbf{Max.} & \textbf{Kurtosis} \\\hline\hline
		\multicolumn{7}{c}{\textbf{Europe}} \\
		\textbf{\begin{tabular}[c]{@{}l@{}}Fertilizer (N)\\ ($gN/m^2$)\end{tabular}} & 11.2046 & 9.59 & 6.1996 & 0 & 44.48 & 3.5058 \\
		\begin{tabular}[c]{@{}l@{}}\textbf{GDP}\\ (Const. 2011 \\ PPP \$/1000)\end{tabular} & 34.5437 & 34.8872 & 10.8505 & 10.0695 & 97.8642 & 1.1158 \\
		\textbf{Dryness} & 0.2783 & 0 & 0.452 & 0 & 2.7099 & 2.7927 \\
		\textbf{Wetness} & 0.4361 & 0.1662 & 0.5423 & 0 & 2.6648 & 0.3427 \\
		\textbf{Price Agr.} & 82.9964 & 82.79 & 16.9829 & 20.33 & 148.13 & 1.6585 \\\hline
		\multicolumn{7}{c}{\textbf{South America}} \\
		\textbf{\begin{tabular}[c]{@{}l@{}}Fertilizer (N)\\ ($gN/m^2$)\end{tabular}} & 3.8203 & 3.21 & 2.7163 & 0 & 21.93 & 7.546 \\
		\begin{tabular}[c]{@{}l@{}}\textbf{GDP}\\ (Const. 2011 \\ PPP \$/1000)\end{tabular} & 11.9857 & 11.56 & 3.4558 & 3.4687 & 21.9983 & 0.0227 \\
		\textbf{Dryness} & 0.2778 & 0 & 0.447 & 0 & 3.6238 & 2.6907 \\
		\textbf{Wetness} & 0.4273 & 0.1384 & 0.5555 & 0 & 3.265 & 0.889 \\
		\textbf{Price Agr.} & 41.4037 & 37.31 & 24.4936 & -1.3397 & 113.28 & -0.5323 \\\hline
		\multicolumn{7}{c}{\textbf{South East Asia}} \\
		\textbf{\begin{tabular}[c]{@{}l@{}}Fertilizer (N)\\ ($gN/m^2$)\end{tabular}} & 7.588 & 7.15 & 4.1817 & 0 & 39.73 & 2.9979 \\
		\begin{tabular}[c]{@{}l@{}}\textbf{GDP}\\ (Const. 2011 \\ PPP \$/1000)\end{tabular} & 5.173 & 4.0498 & 3.7942 & 1.015 & 23.2242 & 5.0619 \\
		\textbf{Dryness} & 0.3154 & 0 & 0.4737 & 0 & 2.6268 & 1.8175 \\
		\textbf{Wetness} & 0.372 & 0.0601 & 0.5041 & 0 & 2.874 & 1.1683 \\
		\textbf{Price Agr.} & 47.4911 & 40.79 & 25.0958 & 5.84 & 135.86 & 0.0253 \\\hline
		\multicolumn{7}{c}{\textbf{Africa}} \\
		\textbf{\begin{tabular}[c]{@{}l@{}}Fertilizer (N)\\ ($gN/m^2$)\end{tabular}} & 1.1405 & 0.28 & 3.9961 & 0 & 49.25 & 61.3191 \\
		\begin{tabular}[c]{@{}l@{}}\textbf{GDP}\\ (Const. 2011 \\ PPP \$/1000)\end{tabular} & 3.0984 & 1.5285 & 3.5376 & 0.3705 & 40.0158 & 7.6428 \\
		\textbf{Dryness} & 0.3901 & 0.1416 & 0.5085 & 0 & 3.4016 & 1.5376 \\
		\textbf{Wetness} & 0.2527 & 0 & 0.4262 & 0 & 2.9603 & 3.693 \\
		\textbf{Price Agr.} & 54.4285 & 50.69 & 28.8513 & -0.245 & 143.59 & -0.5468 \\\hline\hline
	\end{tabular}}\\
\footnotesize{Note: Dryness and Wetness are computed as the number of standard deviations around the mean with dryness\\ taking negative values. We considered the absolute value of dryness in order to ease its interpretation.}
	\label{tab:tab1}
\end{table}
\begin{table}[ht]
	\centering
	\caption{Countries in Each Macro--region with the Relative Number of Cells (Units) per Year.}	
	\begin{tabular}{lclclc}
		\hline\hline
		\multicolumn{1}{c}{\textbf{Country}} & \textbf{Cells} & \multicolumn{1}{c}{\textbf{Country}} & \textbf{Cells} & \multicolumn{1}{c}{\textbf{Country}} & \textbf{Cells} \\ \hline\hline
		\multicolumn{4}{c}{\textbf{Europe}} & \multicolumn{2}{c}{\textbf{South East Asia}} \\ 
		Austria & 55 & Finland & \multicolumn{1}{c|}{146} & Malaysia & 113 \\
		Norway & 164 & France & \multicolumn{1}{c|}{278} & Bangladesh & 33 \\
		Czech Rep. & 61 & Sweden & \multicolumn{1}{c|}{111} & Lao PDR & 47 \\
		Ireland & 35 & Spain & \multicolumn{1}{c|}{202} & India & 1145 \\
		Luxembourg & 1 & Belgium & \multicolumn{1}{c|}{25} & Indonesia & 537 \\
		Netherlands & 12 & Switzerland & \multicolumn{1}{c|}{19} & Pakistan & 273 \\
		Germany & 164 & Italy & \multicolumn{1}{c|}{134} & Vietnam & 132 \\
		Denmark & 31 & Hungary & \multicolumn{1}{c|}{60} & Philippines & 108 \\
		United Kingdom & 149 & Poland & \multicolumn{1}{c|}{165} & Nepal & 39 \\
		Portugal & 50 &  & \multicolumn{1}{l|}{} & Cambodia & 65 \\
		Greece & 66 &  & \multicolumn{1}{l|}{} & Thailand & 211 \\ \hline
		\multicolumn{4}{c}{\textbf{Africa}} & \multicolumn{2}{c}{\textbf{South America}} \\
		Guinea & 77 & South Africa & \multicolumn{1}{c|}{294} & Colombia & 181 \\
		Botswana & 32 & Cameroon & \multicolumn{1}{c|}{107} & Suriname & 5 \\
		Dem. Rep. Congo & 600 & Niger & \multicolumn{1}{c|}{129} & Guyana & 22 \\
		Morocco & 103 & Nigeria & \multicolumn{1}{c|}{297} & Paraguay & 128 \\
		Mali & 165 & Algeria & \multicolumn{1}{c|}{96} & Argentina & 509 \\
		Mozambique & 250 & Eq. Guinea & \multicolumn{1}{c|}{8} & Peru & 207 \\
		Namibia & 50 & Togo & \multicolumn{1}{c|}{7} & Uruguay & 59 \\
		Tanzania & 290 & Egypt & \multicolumn{1}{c|}{58} & Ecuador & 60 \\
		Malawi & 38 & Rwanda & \multicolumn{1}{c|}{7} & Brazil & 1921 \\
		Cote d'Ivoire & 89 & Burundi & \multicolumn{1}{c|}{3} & Bolivia & 192 \\
		Ghana & 86 & Madagascar & \multicolumn{1}{c|}{178} & Chile & 78 \\
		Tunisia & 51 & Burkina Faso & \multicolumn{1}{c|}{83} & Venezuela & 146 \\
		Senegal & 68 & Kenya & \multicolumn{1}{c|}{104} &  & \multicolumn{1}{l}{} \\
		Ethiopia & 223 &  & \multicolumn{1}{l|}{} &  & \multicolumn{1}{l}{}\\\hline\hline
	\end{tabular}
	\label{tab:tab2}
\end{table}
\begin{table}[ht]
	\centering
	\caption{Estimated coefficients through QMLE of model in equation \eqref{eq:timediff}.}
	\resizebox{0.80\textwidth}{0.47\textheight}{
	\begin{tabular}{lcccc}
		\hline\hline
		& \textbf{Europe} & \textbf{\begin{tabular}[c]{@{}c@{}}South \\ America\end{tabular}} & \textbf{\begin{tabular}[c]{@{}c@{}}South East\\ Asia\end{tabular}} & \textbf{Africa} \\
		\hline\hline
		$\phi$ & -2.49e-02*** & -1.14e-01*** & -2.07e-01*** & -1.05e-01*** \\
		& (-5.59e-03) & (0.00439) & (0.00450) & (0.00395) \\
		$\gamma$ & 0.01575* & 0.10461*** & 0.159*** & 0.0553*** \\
		& (0.00614) & (0.00503) & (0.00508) & (0.00464) \\
		$GDP_t$ & 0.04968*** & 0.05981** & -3.41e-02* & 0.00308 \\
		& (0.01154) & (0.01931) & (0.01660) & (0.01100) \\
		$GDP_t^2$ & -5.00e-04*** & 0.00018 & -5.00e-05 & -1.90e-04 \\
		& (0.00012) & (0.00062) & (0.00053) & (0.00025) \\
		$DRY_t$ & 0.00635 & 0.01277*** & -4.78e-03 & -1.40e-03 \\
		& (0.00567) & (0.00372) & (0.00368) & (0.00238) \\
		$WET_t$ & 0.01301** & -1.01e-02*** & 0.0021 & -1.30e-03 \\
		& (0.00479) & (0.00279) & (0.00315) & (0.00264) \\
		$PAO_t$ & 0.00111** & 0.0016*** & -1.20e-03*** & 0.00024+ \\
		& (0.00039) & (0.00041) & (0.00034) & (0.00013) \\
		$GDP_{t-1}$ & -2.68e-02*** & -1.92e-02*** & 0.02880*** & 0.0267*** \\
		& (0.00547) & (0.00474) & (0.00753) & (0.00672) \\
		$DRY_{t-1}$ & -4.76e-02** & -2.69e-03 & -3.03e-03 & -2.66e-03 \\
		& (0.01815) & (0.01120) & (0.00575) & (0.00292) \\
		$WET_{t-1}$ & 0.02183 & 0.02968** & -2.46e-03 & -2.14e-03 \\
		& (0.01516) & (0.00971) & (0.00502) & (0.00343) \\
		$PAO_{t-1}$ & 0.00078* & -4.41e-03*** & -1.32e-03*** & -3.00e-04* \\
		& (0.00039) & (0.00043) & (0.00035) & (0.00013) \\
		$DRY_{t-1}\times GDP_t$ & 0.00142** & 0.0015+ & 0.00021 & 2.00e-05 \\
		& (0.00052) & (0.00090) & (0.00090) & (0.00060) \\
		$WET_{t-1}\times GDP_t$ & -6.70e-04 & -2.81e-03*** & 0.00032 & -1.11e-03+ \\
		& (0.00041) & (0.00077) & (0.00061) & (0.00062) \\
		$\rho$ & 0.83588*** & 0.77474*** & 0.84480*** & 0.74264*** \\
		& (0.00286) & (0.00268) & (0.00233) & (0.00286) \\
		\hline
		$N\times T$ & 36,632 & 66,652 & 51,357 & 66,367 \\
		$N$ dimension & 1,928 & 3,508 & 2,703 & 3,493 \\
		$T$ dimension & 19 & 19 & 19 & 19 \\
		\hline\hline
		\multicolumn{5}{l}{\small{Standard errors in parenthesis. P--value: + p \textless 0.1, * p \textless 0.05, ** p \textless 0.01, *** p \textless 0.001.}}
	\end{tabular}}
	\label{tab:tab3}
\end{table}
\begin{table}[ht]
	\centering
	\caption{Estimated coefficients through QMLE of model in equation \eqref{eq:sarar}, including the price of fertilizer (PF).}
	\resizebox{0.78\textwidth}{0.47\textheight}{
		\begin{tabular}{lcccc}
			\hline\hline
			& \textbf{Europe} & \textbf{\begin{tabular}[c]{@{}c@{}}South\\ America\end{tabular}} & \textbf{\begin{tabular}[c]{@{}c@{}}South East\\ Asia\end{tabular}} & \textbf{Africa} \\ 
			\hline\hline
			$\phi$ & -0.12415*** & -0.10765*** & -0.201*** & 0.01792* \\
			& (0.00861) & (0.00621) & (0.00677) & (0.00758) \\
			$\gamma$ & 0.09785*** & 0.04483*** & 0.183*** & -0.04176** \\
			& (0.00953) & (0.01248) & (0.00762) & (0.01398) \\
			$GDP_t$ & 0.02498 & 0.7732*** & 0.03810 & -0.10846 \\
			& (0.01605) & (0.15851) & (0.03270) & (0.08665) \\
			$GDP_t^2$ & -0.00025 & -0.01565*** & -0.00088 & 0.00386 \\
			& (0.00018) & (0.00442) & (0.00083) & (0.00941) \\
			$DRY_t$ & -0.0015 & 0.04484*** & -0.00543 & -0.01082* \\
			& (0.00894) & (0.01347) & (0.00573) & (0.00442) \\
			$WET_t$ & -0.00219 & -0.02179+ & 0.00959* & 0.01117* \\
			& (0.00764) & (0.01175) & (0.00438) & (0.00520) \\
			$PAO_t$ & 0.00055 & 0.00055 & -0.00202*** & 0.00023 \\
			& (0.00064) & (0.00344) & (0.00044) & (0.00042) \\
			$PF_t$ & 0.02719 & 0.2874 & -0.0278*** & -0.0071 \\
			& (0.04247) & (0.18078) & (0.00603) & (0.00632) \\
			$GDP_{t-1}$ & -0.03223*** & -0.14543*** & 0.00347 & 0.10063* \\
			& (-0.0074) & (0.03346) & (0.0132) & (0.04007) \\
			$DRY_{t-1}$ & -0.06211* & 0.03106 & 0.00351 & -0.00395 \\
			& (0.0285) & (0.03812) & (0.00924) & (0.00703) \\
			$WET_{t-1}$ & 0.05518* & 0.05116 & -0.00725 & 0.00114 \\
			& (0.0232) & (0.03337) & (0.00741) & (0.00835) \\
			$PAO_{t-1}$ & 0.00110+ & 0.00540+ & 0.00e+00 & -0.00094+ \\
			& (0.00066) & (0.00319) & (0.00046) & (0.00048) \\
			$PF_{t-1}$ & 0.05620 & -0.06559 & -0.020700*** & -0.00138 \\
			& (0.04241) & (0.07353) & (0.00538) & (0.00831) \\
	    	        $DRY_{t-1}\times GDP_t$ & 0.00180* & -0.00034 & -0.00157 & 0.00092 \\
			& (0.00080) & (0.00280) & (0.00133) & (0.00104) \\
			$DRY_{t-1}\times GDP_t$ & -0.00161** & -0.00497* & 0.00071 & 0.00120 \\
			& (0.00061) & (0.00253) & (0.00078) & (0.00125) \\			
                        $\rho$ & 0.83086*** & 0.14275*** & 0.84345*** & 0.25700*** \\
			& (0.0043) & (0.00888) & (0.00325) & (0.01039) \\			
			$\lambda$ & -- & 0.84802*** & -- & 0.95164*** \\ 
			& -- & (0.00599) & -- & (0.00294) \\
			\hline
			$N\times T$ & 16,580 & 32,070 & 26,170 & 19,100 \\
			$N$ dimension & 1,658 & 3,207 & 2,617 & 1,910 \\
			$T$ dimension & 10 & 10 & 10 & 10 \\
			\hline\hline
			\multicolumn{5}{l}{\small{Standard errors in parenthesis. P-value: + p \textless 0.1, * p \textless 0.05, ** p \textless 0.01, *** p \textless 0.001.}}
		\end{tabular}}
	\label{tab:tab3c}
\end{table}
\begin{table}[ht!]
	\centering
	\caption{Time--invariant Marginal Effects from SAR and EC models in eq. \eqref{eq:timediff} and \eqref{eq:ecm}, respectively.}
	\resizebox{0.8\textwidth}{0.48\textheight}{
	\begin{tabular}{clcccc}
		\hline\hline
\multicolumn{6}{c}{\textit{\textbf{SDPD Model}}} \\ \hline\hline
		\textbf{Macro--area} & \multicolumn{1}{c}{\textbf{Variable}} & \textbf{\begin{tabular}[c]{@{}c@{}}Effect\end{tabular}} & \textbf{Direct} & \textbf{Indirect} & \textbf{Total} \\ \hline
\multirow{4}{*}[-1cm]{Europe} & \multirow{2}{*}[-0.4cm]{$DRY_{t}$} & \begin{tabular}[c]{@{}c@{}}short\end{tabular} & 0.008805291 & 1.741442e-05 & 0.008822705 \\
		&  & \begin{tabular}[c]{@{}c@{}}long\end{tabular} & 0.008521506 & 1.630236e-05 & 0.008537809 \\ \cline{3-6} 
		& \multirow{2}{*}[-0.4cm]{$WET_{t}$} & \begin{tabular}[c]{@{}c@{}}short\end{tabular} & 0.018056484 & 3.571070e-05 & 0.018092195 \\
		&  & \begin{tabular}[c]{@{}c@{}}long\end{tabular} & 0.017474544 & 3.343028e-05 & 0.017507974 \\ \cline{3-6} 
		& \multirow{2}{*}[-0.4cm]{$PAO_{t}$} & \begin{tabular}[c]{@{}c@{}}short\end{tabular} & 0.001537706 & 3.041156e-06 & 0.001540748 \\
		&  & \begin{tabular}[c]{@{}c@{}}long\end{tabular} & 0.001488148 & 2.846953e-06 & 0.001490995 \\ \hline
\multirow{4}{*}[-0.8cm]{\begin{tabular}[c]{@{}c@{}}South\\ America\end{tabular}} & \multirow{2}{*}[-0.4cm]{$DRY_{t}$} & \begin{tabular}[c]{@{}c@{}}short\end{tabular} & 0.016041268 & 1.158859e-05 &  0.016052856 \\
		&  & \begin{tabular}[c]{@{}c@{}}long\end{tabular} & 0.014605642 & 1.133174e-05 & 0.014616974 \\ \cline{3-6} 
		& \multirow{2}{*}[-0.4cm]{$WET_{t}$} & \begin{tabular}[c]{@{}c@{}}short\end{tabular} & -0.012714951 & -9.185577e-06 & -0.012724137 \\
		&  & \begin{tabular}[c]{@{}c@{}}long\end{tabular} & -0.011577017 & -8.981987e-06 & -0.011585999 \\ \cline{3-6} 
                & \multirow{2}{*}[-0.4cm]{$PAO_{t}$} & \begin{tabular}[c]{@{}c@{}}short\end{tabular} & 0.002005386 & 1.448737e-06 & 0.002006834 \\
		&  & \begin{tabular}[c]{@{}c@{}}long\end{tabular} & 0.001825912 & 1.416627e-06 & 0.001827329 \\ \hline
\multirow{4}{*}[-0.8cm]{\begin{tabular}[c]{@{}c@{}}South--East\\ Asia\end{tabular}} & \multirow{2}{*}[-0.4cm]{$DRY_{t}$} & \begin{tabular}[c]{@{}c@{}}short\end{tabular} & -0.006582302 & -8.971406e-06 & -0.006591273 \\
		&  & \begin{tabular}[c]{@{}c@{}}long\end{tabular} & -0.005349902 & -6.731993e-06 & -0.005356634 \\ \cline{3-6} 
		& \multirow{2}{*}[-0.4cm]{$WET_{t}$} & \begin{tabular}[c]{@{}c@{}}short\end{tabular} & 0.002888857  & 3.937393e-06 & 0.002892794 \\
		&  & \begin{tabular}[c]{@{}c@{}}long\end{tabular} & 0.002347978 & 2.954554e-06 & 0.002350933 \\ \cline{3-6} 
                & \multirow{2}{*}[-0.4cm]{$PAO_{t}$} & \begin{tabular}[c]{@{}c@{}}short\end{tabular} & -0.001650949 & -2.250175e-06 & -0.001653199 \\
		&  & \begin{tabular}[c]{@{}c@{}}long\end{tabular} & -0.001341843 & -1.688493e-06 & -0.001343531 \\ \hline
\multirow{4}{*}[-1cm]{Africa} & \multirow{2}{*}[-0.4cm]{$DRY_{t}$} & \begin{tabular}[c]{@{}c@{}}short\end{tabular} & -0.0017101155 & -1.064206e-06 & -0.0017111797 \\
		&  & \begin{tabular}[c]{@{}c@{}}long\end{tabular} & -0.0015224597 & -8.673616e-07 & -0.0015233271 \\ \cline{3-6} 
		& \multirow{2}{*}[-0.4cm]{$WET_{t}$} & \begin{tabular}[c]{@{}c@{}}short\end{tabular} & -0.0015950535 & -9.926028e-07 & -0.0015960461 \\
		&  & \begin{tabular}[c]{@{}c@{}}long\end{tabular} & -0.0014200238 & -8.090028e-07 & -0.0014208328 \\ \cline{3-6} 
                & \multirow{2}{*}[-0.4cm]{$PAO_{t}$} & \begin{tabular}[c]{@{}c@{}}short\end{tabular} & 0.0002893142 & 1.800404e-07 & 0.0002894943 \\
		&  & \begin{tabular}[c]{@{}c@{}}long\end{tabular} & 0.0002575670 & 1.467387e-07 & 0.0002577137 \\ \hline
\multicolumn{6}{c}{\textit{\textbf{EC Model}}} \\ \hline
\textbf{Macro--area} & \multicolumn{1}{c}{\textbf{Variable}} & \textbf{\begin{tabular}[c]{@{}c@{}}Effect\end{tabular}} & \textbf{Direct} & \textbf{Indirect} & \textbf{Total} \\ \hline
		Europa & \multicolumn{1}{c}{$\by_{t-1}$} & / & -1.816695 & -0.005607266 & -1.822302 \\
		\begin{tabular}[c]{@{}c@{}}South America\end{tabular} & \multicolumn{1}{c}{$\by_{t-1}$} & / & -1.69087 & -0.002041492 & -1.692912 \\
		\begin{tabular}[c]{@{}c@{}}South--East Asia\end{tabular} & \multicolumn{1}{c}{$\by_{t-1}$} & / & -2.106944 & -0.004490682 & -2.111435 \\
		Africa & \multicolumn{1}{c}{$\by_{t-1}$} & / & -1.594045 & -0.00166058 & -1.595706 \\ \hline\hline
	\end{tabular}}
\label{tab:tab5}
\end{table}
\clearpage
%
%
\section{Figures}
\end{document}